\documentclass{cpbtex}

\usepackage{graphicx}% Include figure files

\begin{document}

\title{Recent advances in Ultralong-range Rydberg molecules}	

\author{Jingxu Bai(白景旭)$^{1,2}$, \ Yuechun Jiao(焦月春)$^{1,2}$, \ Xiao-Qiang Shao(邵晓强)$^{3,4}$, \\ Weibin Li(李伟斌)$^{5}$\thanks{Corresponding author. E-mail:~weibin.li@nottingham.ac.uk}, \ and \ Jianming Zhao(赵建明)$^{1,2}$\thanks{Corresponding author. E-mail:~jmzhao@sxu.edu.cn}\\
$^{1}${State Key Laboratory of Quantum Optics Technologies and Devices,}\\%
{Institute of Laser Spectroscopy, Shanxi University, Taiyuan 030006, China}\\% 
$^{2}${Collaborative Innovation Center of Extreme Optics,}\\%
{Shanxi University, Taiyuan 030006, China}\\ %
$^{3}${Center for Quantum Science and School of Physics, }\\%
{Northeast Normal University, Changchun, Jilin 130024, China}\\ %
$^{4}${Institute of Quantum Science and Technology, }\\%
{Yanbian University, Yanji 133002, China}\\ %
$^{5}${School of Physics and Astronomy and Centre for the}\\ {Mathematics and Theoretical Physics of Quantum Non-equilibrium Systems, }\\%
{University of Nottingham, Nottingham NG7 2RD, United Kingdom}}
		
\date{\today}
\maketitle

\begin{abstract}
Rydberg molecule, formed by one or more Rydberg atoms, exhibits remarkable properties, including an exceptionally large spatial extent, rich rovibrational level structures, permanent electric dipole moments, and a pronounced sensitivity to external fields. Based on the underlying binding mechanisms, Rydberg molecules can be divided into three categories, the ground-Rydberg molecule that is bound via a low-energy electron-atom scattering interaction between ground atom and Rydberg electron, the Rydberg-Rydberg molecule that is bound via a long-range electrostatic interaction between Rydberg atoms, and the ion-Rydberg molecule that is bound via single- or multi-polar interactions between Rydberg atom and ion. This review focuses on recent theoretical and experimental advances in diatomic Rydberg molecules, covering their formation and binding mechanisms, potential energy curves, experimental observations, and spectroscopic properties, with the aim of providing a comprehensive overview of the current state and future prospects of this rapidly developing field.
\end{abstract}
		
		

\section{Introduction}\label{sec1}

\begin{figure}[htbp]
\begin{center}
\includegraphics[width=0.9\textwidth]{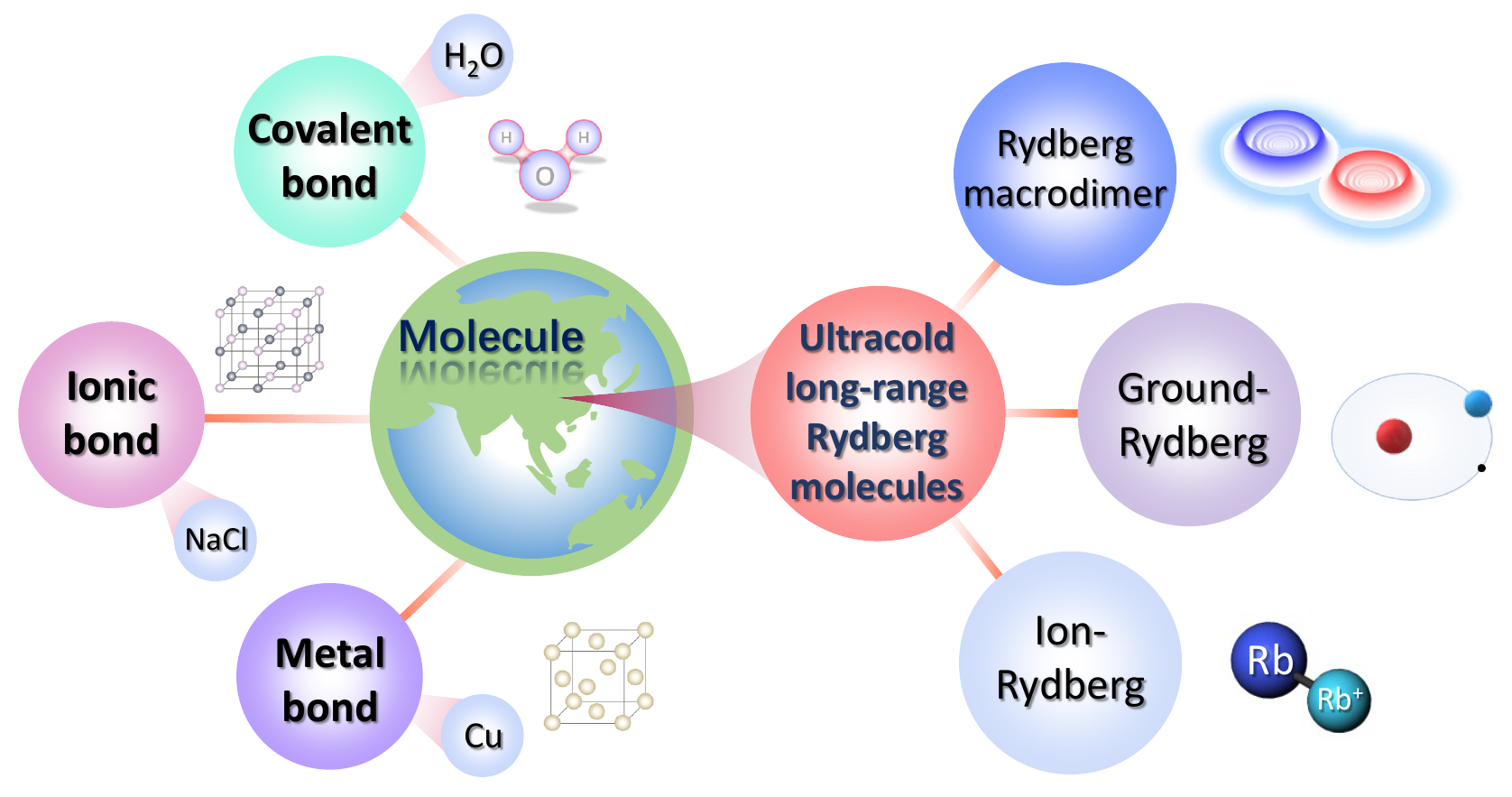}
\end{center}
\caption{ Different molecule types. Molecules in nature are mainly bound by ionic bonds, covalent bonds and metallic bonds. The new type of long-range Rydberg molecules can be divided into three categories, ground-Rydberg molecules, Rydberg macrodimer, and ion-Rydberg molecules. }\label{Fig1}
\end{figure}

Molecules constitute the fundamental building blocks of matter in nature and arise from interactions between atoms through various types of chemical bonding~\cite{Pauling1960,Ashcroft1976,Shriver2014}. As schematically illustrated in Figure~\ref{Fig1}, the vast majority of molecules encountered in both natural and chemical systems are stabilized by finite-range interactions, primarily covalent bonds (e.g., $\text{H}_2\text{O}$), ionic bonds (e.g., NaCl), and metallic bonds (e.g., Cu). These conventional molecules have internuclear separations on the angstrom scale and binding energies dictated by the overlap of the valence electron wavefunctions. This electronic configuration fundamentally underlies all chemical reactions, biological processes, and material properties.

Beyond these traditional molecular species, advances in atomic and molecular physics have revealed that atoms can also form molecules through unconventional binding mechanisms that fall outside the standard paradigm of chemical bonding~\cite{Blaney1976,Chin2010}. The realization of these exotic molecular systems has been enabled by laser cooling and trapping techniques~\cite{Metcalf1999,Chu1985}, which suppress thermal motion at ultracold temperatures and allow precise control of interatomic interactions, thereby making it possible to form fragile and highly extended molecular states that are inaccessible under ordinary conditions~\cite{Carr2009,Saffman2010}.

A particularly important role in this context is played by Rydberg atoms~\cite{gallagher1994}, which are atoms excited to high principal quantum numbers. Rydberg atoms exhibit pronounced properties, including extremely large orbital radii, long lifetimes, and strong sensitivity to external electric and magnetic fields. Their valence electrons can extend over hundreds or even thousands of Bohr radii, resulting in enhanced dipole moments and strong long-range interactions. These unique characteristics make Rydberg atoms an ideal platform for exploring strong interactions~\cite{Lukin2001,Dudin2012,Vogt2006,shao2024rydberg}, many-body and ultrafast physics~\cite{Olmos2009,Takei2016,Mizoguchi2020,Browaeys2020,Wenbin2021}, quantum optics~\cite{Fleischhauer2005}, quantum information~\cite{Li2013,Baur2014,Gorniaczyk2016}, quantum computation and quantum simulation~\cite{Saffman2010,Weimer2010,Lukin2001,Wu2021} and sensing~\cite{sedlacekMicrowaveElectrometryRydberg2012,jingAtomicSuperheterodyneReceiver2020,yuanQuantumSensingMicrowave2023}.

When Rydberg atoms interact with perturbers, such as ground-state atoms, Rydberg atoms, or ions, a remarkable class of exotic molecules, known as Rydberg molecules~\cite{Shaffer2018,Fey2019,Dunning2024}, can be formed. The underlying binding mechanisms are fundamentally different from those of conventional molecules. As depicted in Fig.~\ref{Fig1}, based on the binding mechanisms, long-range Rydberg molecules are generally classified into three categories: ground–Rydberg molecules formed by the low-energy electrons scattering~\cite{Greene2000,Hamilton2002}, Rydberg macrodimers formed by the long-range multipole interactions~\cite{Boisseau2002}, and ion-Rydberg molecules formed by monopole and multipole interactions~\cite{Duspayev2021Long,Dei2021}. Owing to their peculiar electronic structures and extended spatial scales, Rydberg molecules exhibit properties distinct from those of both ionic and covalent molecules. These unconventional interactions result in molecular states with extraordinarily large bond lengths, ultralow binding energies, rich rovibrational structures, permanent electric dipole moments, and a pronounced sensitivity to external fields.

The study of Rydberg molecules therefore lies at the intersection of atomic physics, molecular physics, and quantum optics, significantly extending the traditional boundaries of molecular science. Owing to their long-range and highly tunable interactions, Rydberg molecules provide a versatile platform for investigating the microscopic mechanisms of molecular formation~\cite{Guttridge2025} and interaction and exploring quantum spatial correlations~\cite{Weimer2010,Manthey2015,Whalen2019MP,Whalen2019PRA}, few- and many-body physics~\cite{Schmidt2016,Schlagmuller2016,Camargo2018,Kleinbach2018} in a controlled setting. Moreover, their exceptional response to external fields and strong interparticle interactions make them promising candidates for quantum information processing applications ~\cite{Lukin2001,DeMille2002,Rabl2006}. The related spectroscopic techniques enable the probing of the molecular structure and intrinsic properties, offering direct insight into their underlying binding mechanisms.

This review provides a comprehensive overview of the historical development and recent progress in the study of long-range diatomic Rydberg molecules. We focus on the distinct binding mechanisms and summarize both theoretical and experimental advances, including the molecular modeling, the high-resolution photoassociation spectroscopy, and the characterization of different classes of Rydberg molecular states. Finally, we discuss the current challenges and outline future research directions in this rapidly evolving field.

\section{Ground-Rydberg molecule}\label{sec2}

The ground-Rydberg molecule is the most extensively studied type of ultralong-range Rydberg molecule (ULRM), wherein the ground-state atom is bound to the Rydberg electron orbit through the scattering interaction between the Rydberg electron and the ground-state atom~\cite{Greene2000,Hamilton2002}. The ground-Rydberg molecule exhibits a bond length of several thousand Bohr radii, a scale comparable to the size of a bacterium. The binding potential energy is determined by the probability density of the Rydberg wave function and the energy-dependent scattering length between the two particles. These Rydberg molecules have a large permanent dipolar moment of several thousand $ea_0$~\cite{Li2011,booth2015,Niederpruem2016,baiPRR2020,jiao2023}. The ground-state Rydberg molecules provide an important platform for understanding ultracold low-energy atomic collisions~\cite{Anderson2014,Sa2015,Boettcher2016,Maclennan2019,Engel2019,Wang2024}, external regulation of molecular properties~\cite{Lesanovsky2006,krupp2014,Kurz2013,Gaj2015,Niederpruem2016,Hummel2018,Hummel2019}, verifying scattering phase shift and negative ion resonance~\cite{Fabrikant1986,Bahrim2000,Bahrim2001a,Bahrim2001b}, and measuring the DNA strand breaks caused by negative ion resonance~\cite{Bald2006,Simons2006,Martin2004,Caron2003,Alizadeh2015}.

\subsection{Binding mechanism }\label{sec2.1} 

The binding mechanism of ground-Rydberg molecules is based on a low-energy electron-atom scattering interaction between the ground-state atom and the Rydberg electron with a negative electron-atom-wave scattering length. 
This low-energy scattering interaction gives rise to a substantial potential well that confines the ground-state atom within the Rydberg orbit, thereby forming a stable Rydberg molecule~\cite{Greene2000,Hamilton2002}. This low-energy scattering interaction was first observed in 1934 through the pressure broadening phenomenon caused by collision interactions between Rydberg atoms in a vapor cell~\cite{Amaldi1934}, and was subsequently explained by Fermi based on relativistic quantum mechanics, leading to the establishment of pseudopotential theory~\cite{Fermi1934}.

\begin{figure}[htbp]
\begin{center}
\includegraphics[width=0.5\textwidth]{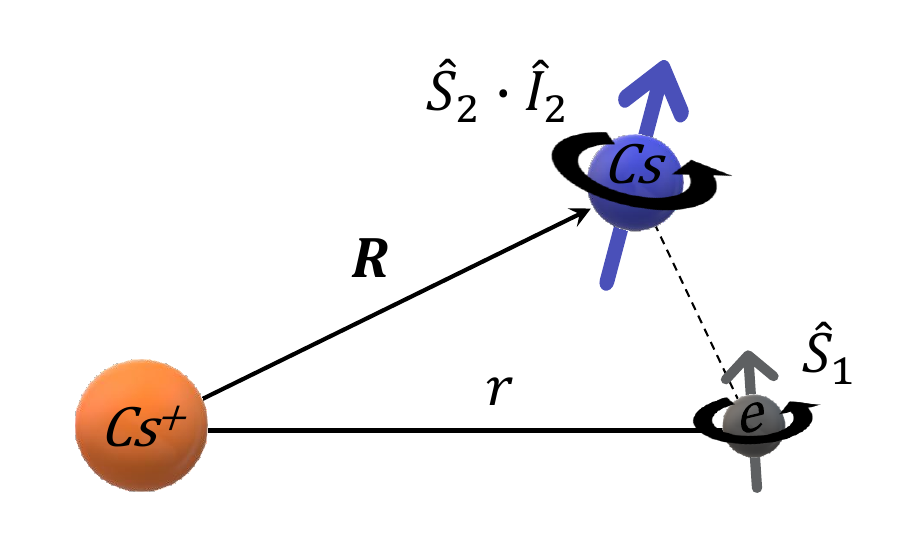}
\end{center}
\caption{ Model of a ground-Rydberg molecule formed by two Cs atoms in the ground-state and Rydberg state. }\label{Fig2}
\end{figure}  

Based on the Fermi pseudopotential theory~\cite{Fermi1934,Omont1977a} and the Born-Oppenheimer approximation~\cite{Born1927Zur}, the nuclear motion of Rydberg atoms can be separated from the electronic motion of Rydberg atoms, the steady-state solution of the Schr\"odinger equation yields an interaction potential energy curve. Figure~\ref{Fig2} shows the theoretical model of the ground-Rydberg molecule. Considering a diatomic system consisting of a ground-state atom and a Rydberg atom, the direction of the axial vector between the Rydberg atomic nucleus and the ground-state atom is the direction of the quantized axis. Then, the Hamiltonian, $\hat{H}(r, R)$, of the diatomic system is 
written as~\cite{Anderson2014}:
\begin{equation}
\hat{H}(r, R) = \hat{H}_{0} + \sum_{i = S, T} \hat{V}(r, R) \hat{P}(i) + A_{\mathrm{HFS}} \, \hat{S}_2 \cdot \hat{I}_2,\label{eq1}
\end{equation}
where $r$ and $\mathbf{R}$ represent the position vectors of the Rydberg electron and ground-state atom, respectively; $\hat{H}_{0}$ denotes the Hamiltonian of the unperturbed Rydberg atom, which incorporates the quantum defect and fine structure of the Rydberg atom; $\hat{V}(r, R)$ represents the scattering interaction between the Rydberg electron and the ground-state atom. The projection operator $\hat{P}(i)$ describes the coupling between the Rydberg electron spin $\hat{S}_1$ and the ground-state atom spin $\hat{S}_2$, expressed specifically as $\hat{P}(i) = \hat{S}_1 \cdot \hat{S}_2 + 3/4$ and $\hat{P}(S) = 1 - \hat{P}(T)$, where $i = S, T$ correspond to the spin singlet ($S$) and spin triplet ($T$) states, representing scattering channels with spin couplings of 0 and 1, respectively. The final term in Eq.~\ref{eq1} denotes the hyperfine interaction between the ground-state atom spin $\hat{S}_2$ and the nuclear spin $\hat{I}_2$, with $A_{\mathrm{HFS}}$ being the hyperfine structure constant.

The scattering interaction between the Rydberg electron
and the ground-state atom is, in the reference frame of the
Rydberg ionic core~\cite{Fermi1934,Omont1977a}:
\begin{equation}
\hat{V}(r, R) = 2\pi \alpha_s(k) \, \delta^{3}(r - R\hat{z})
+ 6\pi [\alpha_p(k)]^{3} \, \delta^{3}(r - R\hat{z}) \, \overleftarrow{\nabla} \cdot \overrightarrow{\nabla},\label{eq2}
\end{equation}
where $\alpha_l(k)$ are the scattering lengths, $k$ is the electron momentum, and $l$ is the scattering partial-wave order ($l=0 $ or 1 for $s$ wave or $p$ wave, respectively).  Within the classically allowed region $r$ for the electron, the semi-classical expression for the momentum of the Rydberg electron is $k(r) = \sqrt{2E_{\mathrm{kin}}} = \sqrt{2(E_{nl} + 1/r)}$, where $E_{nl} = -\tfrac{1}{2}(n - \delta_{nl})^{2}$ is the energy of the Rydberg level.

Scattering length $\alpha_l(k)$ is an important parameter, representing the low-energy scattering interaction between Rydberg electrons and ground-state atoms. In the non-relativistic approach, we use finite-range model potentials for the low-energy electron scattering, and related scattering lengths a $a_S^T(k)$ and $a_{p,J}^T(k)$ are determined by numerically integrating the s-wave singlet and triplet scattering wave functions. 
Employing the improved effective range theory (ERT)~\cite{Blatt1949,OMalley1961}, the relationship between the scattering phase shift and the corresponding scattering length for singlet and triplet wave scattering is then derived as follows~\cite{Omont1977a,Hamilton2002}:
\begin{equation}
a_S^T(k) = -\frac{\tan(\delta_S^T(k))}{k},
\quad\quad
a_{p,J}^T(k) = -\frac{\tan(\delta_{p,J}^T(k))}{k^3}.\label{eq3}
\end{equation}

Numerical solution of the Hamiltonian equation (\ref{eq1}) on a grid of $R$ values yields sets of adiabatic potential energy curves of the ground-Rydberg molecule~\cite{Anderson2014,Bendkowsky2009}.  Furthermore, through the molecular Hamiltonian theory~\cite{yang2008,yang2010,yang2012}, the vibration and rotation wave functions of the molecule can be calculated.

\subsection {Potential energy curves (PEC) }\label{sec2.2} 

Greene and his collaborators firstly refined Fermi's scattering theory in 2000, and calculated adiabatic potential energy curves of Rb ground-Rydberg molecules, predicting the existence of ground-Rydberg molecules~\cite{Greene2000}. The probability density of the calculated molecular wave function is similar to the shape of a trilobite fossil, so called the ``trilobite-type" ground-Rydberg molecules~\cite{Greene2000}. They further demonstrated that Rydberg molecules have a large permanent molecular dipole moment, and the properties of such molecules vary significantly depending on the different orbital angular momenta of the Rydberg electrons~\cite{Khuskivadze2002}. This ground-Rydberg state molecules are classified into the following two categories: 1) low-$l$ ground-Rydberg molecules formed by Rydberg atoms in non-degenerate orbital angular momenta and ground state atoms, with binding energy levels of tens of MHz; 2) high-$l$ ground-Rydberg molecules formed by degenerate high angular momentum Rydberg atoms, with binding energies typically at the GHz level. Shortly thereafter, Hamilton et al. predicted another type of polar molecule, distinct from previous findings from $s$-wave scattering interactions, which arises from $p$-wave scattering in electron-atom interactions named  ``butterfly-type" ground-state-Rydberg molecule due to their electron wave function probability density shape as a butterfly~\cite{Hamilton2002,Lesanovsky2006}.

\begin{figure}[htbp]
\begin{center}
\includegraphics[width=0.8\textwidth]{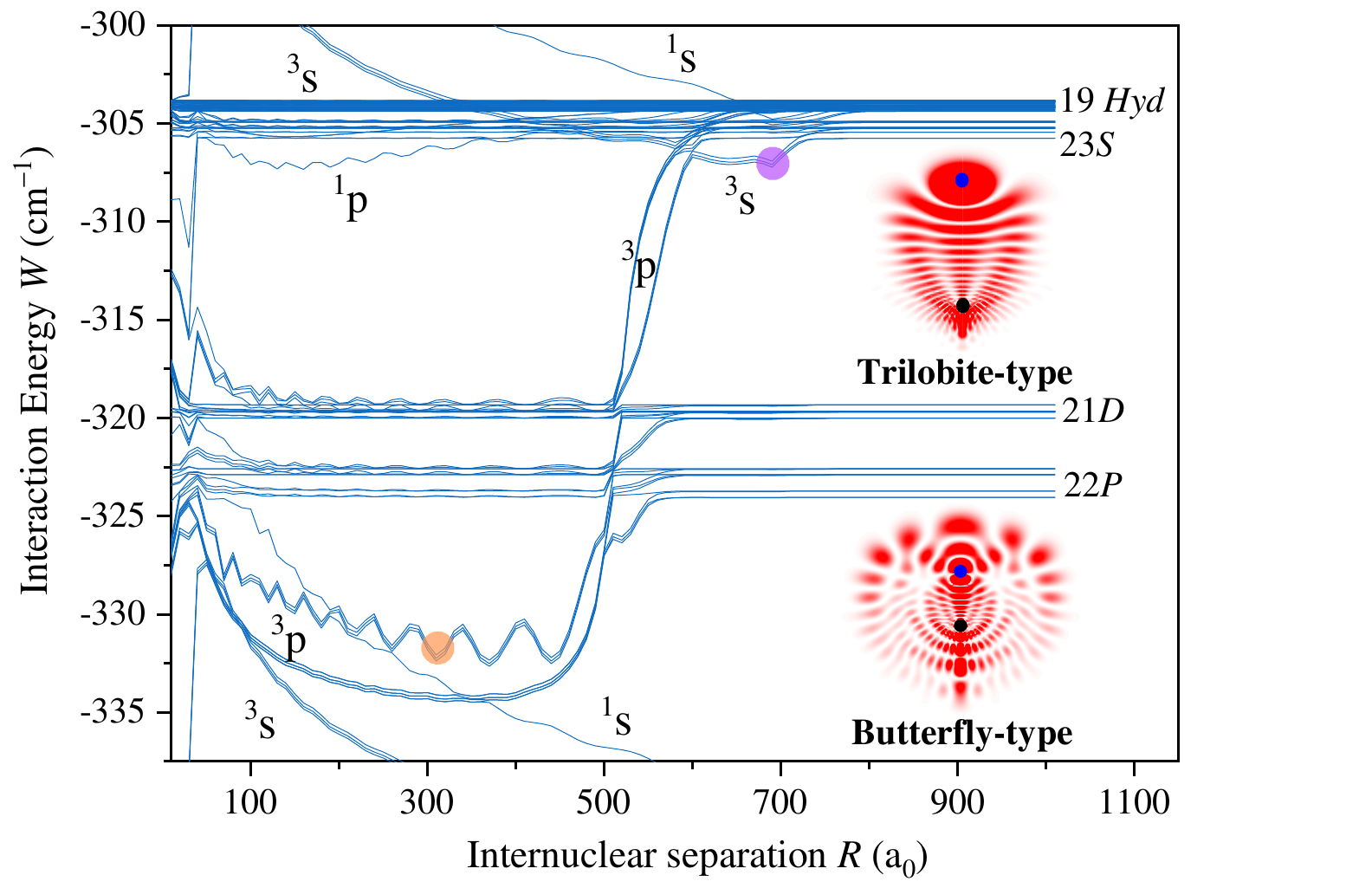}
\end{center}
\caption{Potential energy curves of the Cs molecule near $n = 19$. The inset shows the ``trilobite-type" (``butterfly-type") molecular electron probability density that is formed in the potential wells at the purple (orange) origin markers I (II)~\cite{BaisyPhD2021}. }\label{Fig3}
\end{figure}

Figure~\ref{Fig3} presents the theoretically calculated potential energy curves of Cs ground-Rydberg molecules near $n = 19$. As shown in the figure, a singlet $^1s$-wave scattering gives rise to the repulsive interaction, while the inner potential well induced by singlet $^1p$-wave scattering is counteracted by the triplet $^3p$-wave interaction, which cannot support a stable bound molecular state. In contrast, the potential energy curves arising from triplet $^3s$- and $^3p$-wave scattering exhibit attractive interactions and form bound potential wells, which are capable of supporting stable ground-Rydberg molecules. The outer potential well, located at an internuclear separation of approximately $700~a_0$ (marked as I in the figure), is predominantly generated by the triplet $^3s$-wave scattering. The corresponding electronic probability density exhibits a characteristic three-lobed structure, as illustrated in inset~(I). Molecules bound in this outer well are commonly referred to as ``trilobite-type" Rydberg molecules~\cite{Greene2000}. The inner potential well, appearing at internuclear separations of approximately $200\sim400~a_0$ (marked as II), is mainly formed by triplet $^3p$-wave scattering. The associated electronic probability density displays a butterfly like shape, as shown in inset (II); molecules bound in this well are therefore known as ``butterfly-type" Rydberg molecules~\cite{Hamilton2002}. Compared with the ``trilobite-type" molecules, ``butterfly-type" molecules feature a significantly deeper potential well and a smaller equilibrium internuclear separation. In the electronic wave function illustrations in Figure~\ref{Fig3}, Cs ions are represented by black dots, while ground-state atoms are indicated by blue dots.

It should be noted that the potential well of this ground-Rydberg molecule is influenced by many factors. First, the outermost potential well of the molecular potential energy curve can be well described by $s$-wave scattering interaction, whereas the deep inner potential well at smaller nuclear separations is significantly affected by the $p$-wave scattering interaction because of the higher energy of electrons and the ion nucleus being closer to the Rydberg atom~\cite{baijcp2020}. Second, owing to the spin-orbit coupling interaction of the atom, the fine structure of the Rydberg atom also affects the molecular potential energy curve~\cite{Markus2020}. The effects produced by different seed atoms and Rydberg states are different~\cite{baiPRR2020}. Finally, the hyperfine energy level of the ground-state atom also affects the outermost potential well~\cite{baicpl2020}. Investigations have shown that the shallow potential well formed by the mixed scattering interaction of singlet and triplet states strongly depends on the hyperfine energy level of the ground state atom. The depth of the potential well for the ground-state atom of $F = 3$ is larger than that of $F = 4$; while the deep potential well (triplet) formed by triplet scattering is not affected by the hyperfine interaction of the ground-state atom. After the theoretical predictions, many experiments were conducted to measure the properties of the molecule.

\subsection{Experimental Realization}\label{sec2.3}

\begin{figure}[htbp]
\begin{center}
\includegraphics[width=0.9\textwidth]{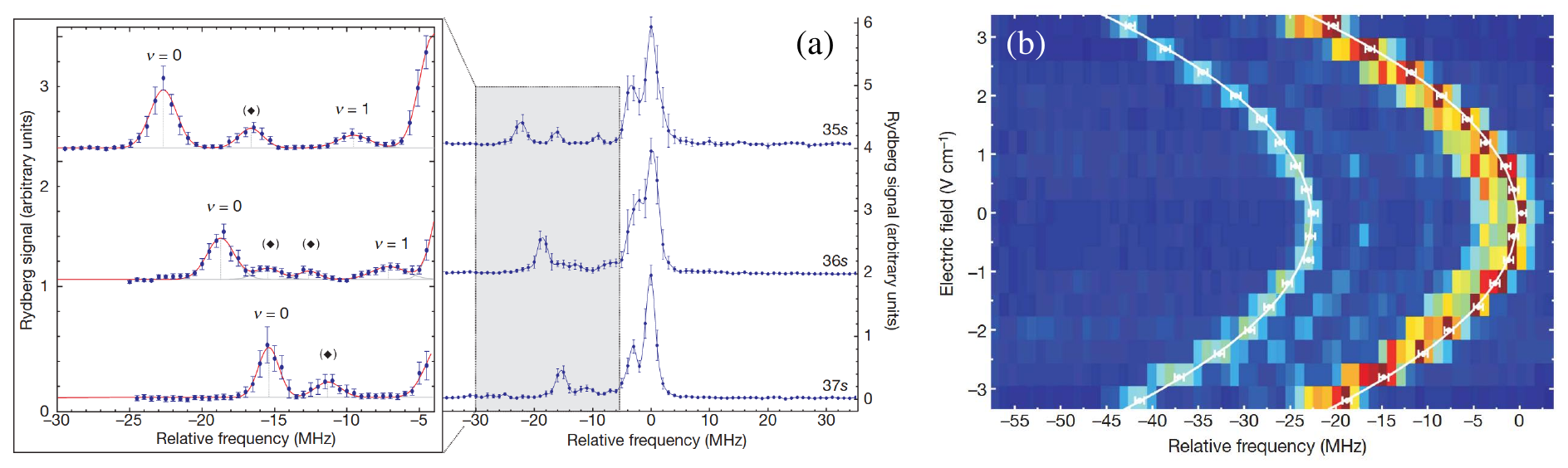}
\end{center}
\caption{ (a) Measured spectra of $35S, 36S$ and $37S$ Rb ground-Rydberg molecules. The right overview is centered on the atomic Rydberg transition. The left side shows the higher resolution molecular lines for $\nu=0,1$ bound state. Diamonds mark the peaks that have not yet been assigned. (b) Stark map of the atomic $35S$ state and the molecular $^3\Sigma(5S–35S)~(\nu=0)$ state, showing an obvious quadratic Stark effect (white lines)~\cite{Bendkowsky2009}. (Reproduced with permission and adapted from Ref.~\cite{Bendkowsky2009}, licensed under a Creative Commons Attribution 4.0 International license.) 
}\label{Fig4}
\end{figure}

In the experiment, the ground-Rydberg molecule was first observed with cold Rb $nS_{1/2}$ ($n=35-37$) state by using a two-photon photo-association scheme~\cite{Bendkowsky2009}, and later with Rb $nP_{1/2,3/2}$ state~\cite{bellos2013} and $nD_{3/2,5/2}$ state~\cite{Anderson2014,krupp2014,Maclennan2019}, as well as with Cs $nS_{1/2}$~\cite{tallant2012,booth2015}, $nP_{3/2}$~\cite{Sa2015} and $nD_{3/2,5/2}$~\cite{baijcp2020,baiPRR2020,baicpl2020,fey2019a} states. Figure~\ref{Fig4}(a) shows the ground-Rydberg molecular spectral lines in the Rb $nS$ states for $n$ = 35, 36, and 37~\cite{Bendkowsky2009}. The left panel displays the enlarged high-resolution spectra marked with the shaded region in the right panel. Both of the vibrational energy levels of $^3\Sigma(5S–nS)~(\nu=0)$ (the leftmost line of each spectrum) and $\nu= 1$ were in good agreement with the model. The small peaks marked as diamonds are unclassified. Based on the observed spectral lines and the binding energies as a function of principal quantum number, the scattering length, $\alpha_{Rb}=-18.5~a_0$, of Rb Rydberg electron and ground atom is derived. They further obtained the Rb ground-Rydberg $nS_{1/2}$ molecules lifetime of $15\sim18~\mu$s, the relative polarizabilities of $\alpha=1,524(4)\times10^7 $~a.u. with the photo-association spectra by changing the time between excitation and field ionization and the different electric fields. Figure~\ref{Fig4}(b) presents the Stark map of the atomic $35S$ state and the molecule $^3\Sigma(5S-35S)~(\nu=0)$ state, spectral peaks of both the atomic state (right) and the molecular state (left) (symbols) show a quadratic Stark effect, agree with the calculations (solid lines). Specially, they found that this kind of molecule has a permanent dipole moment on the order of 1~Debye~\cite{Li2011}. This is the first report of the observation of a permanent electric dipole moment in a homonuclear molecule in which the binding is based on asymmetric electronic excitation between the atoms.

\begin{figure}[htbp]
\begin{center}
\includegraphics[width=\textwidth]{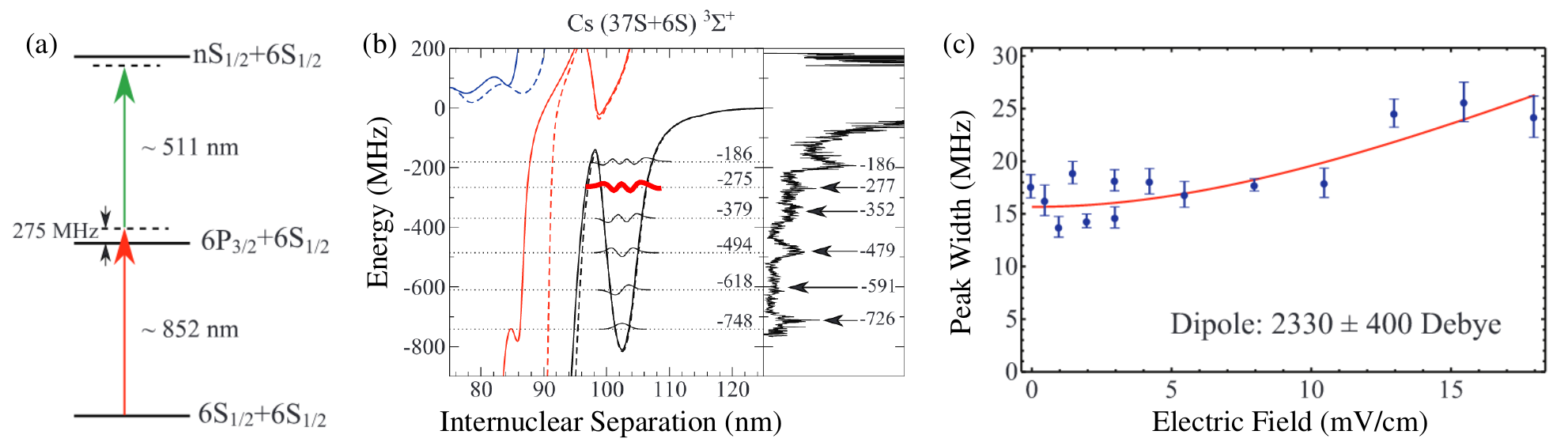}
\end{center}
\caption{ (a) Level diagram for the two-photon excitation scheme. (b) Comparison between the calculated PEC and measured photo-association spectroscopy of Cs ground-Rydberg molecules. Different vibrational levels marked by the dashed lines are clearly distinguishable, and even-parity $(\nu= 0, 2, ..)$ vibrational levels have stronger signals. (c) Linewidth as a function of electric field for the red wavefunction-marked vibrational states in (b), yielding a permanent dipole moment of 2,330(400)~Debye~\cite{booth2015}. (Reproduced with permission and adapted from Ref.~\cite{booth2015}, licensed under a Creative Commons Attribution 4.0 International license.)
}\label{Fig5}
\end{figure}

Furthermore, Cs ``trilobite-type" ground-Rydberg molecules are reported by J. P. Shaffer group~\cite{booth2015}. Using the two-photon association scheme of 
an ultracold cesium atomic ensemble in Figure 5(a), they observed an ultralong-range Cs Rydberg molecules, $^3\Sigma^+(6S-nS)$, with bond lengths of 100~nm and kilo-Debye permanent electric dipole moments. The typical photo-association spectrum for Cs $^3\Sigma^+(6S-37S)$ state is displayed in Figure~\ref{Fig5}(b), where the calculated vibrational levels in the outermost potential well superposed with the associated wave functions (left) and observed spectra (right) are presented. The dashed lines and arrows indicate the different vibrational levels. According to the measured molecular spectra linewidth as a function of the electric field, as shown in Figure~\ref{Fig5}(c) for the $\nu $=5 state marked with the thick red in Fig.~\ref{Fig5}(b), the permanent electric dipole moment of the cesium atom $S$-state ground-Rydberg molecule has a value of more than 2000~Debye, which is 5 orders of magnitude larger than that of water molecules. The measured permanent electric dipole moment is much larger for Cs $(6S-nS)$ molecules than for Rb $(6S-nS)$ molecules because the quantum defect of the cesium atom $S$ state is 4.05, which is close to an integer, thus having a potential energy curve very similar to that of a hydrogen-like atom. This causes the potential energy curve of the ``trilobite-type" molecule with high angular momentum quantum states to intersect with the $S$ state, resulting in the molecule having the characteristics of the cesium atom $S$ state, leading to a huge permanent electric dipole moment.

Regardless of whether in initial experimental studies or theoretical simulations, ground-Rydberg molecules involving $S$ states ($\ell = 0$) possess significant advantages. First, the spherical symmetry of the $S$ state wavefunction allows for a substantial simplification of the theoretical models. Second, the relatively large Franck-Condon factors associated with $S$-state Rydberg molecules lead to higher excitation probabilities in experiments. The $S$-type ground-Rydberg molecules have been widely investigated, and the corresponding molecular parameters, such as the binding energy, lifetime, and permanent dipole moments, have been obtained. Furthermore, one began to study the $P$- and $D$-type ground-Rydberg (Rydberg with $\ell > 0$ ) molecules. $P$- and $D$-type ground-Rydberg molecules have similar molecular spectral structures, forming triplet s-wave scattering $^T\Sigma$ states and singlet and triplet mixed $^{S,T}\Sigma$ molecular states. 

Compared with the $S$-state ground-Rydberg molecules, the investigation of $D$-state ground-Rydberg molecules is substantially more complex. The higher orbital angular momentum gives rise to a strongly anisotropic electronic wavefunction, such that the molecular interaction can no longer be reduced to an effective one-dimensional potential depending only on the internuclear separation, but instead exhibits pronounced orientation dependence and multichannel coupling. In addition, the mixing of higher-partial-wave electron–atom scattering and fine-structure states further fragments the potential energy surfaces and renders the bound states shallow and fragile. These features significantly complicate theoretical modeling and numerical treatment, and at the same time pose major challenges for experimental preparation and spectroscopic identification. 

\begin{figure}[htbp]
\begin{center}
\includegraphics[width=0.8\textwidth]{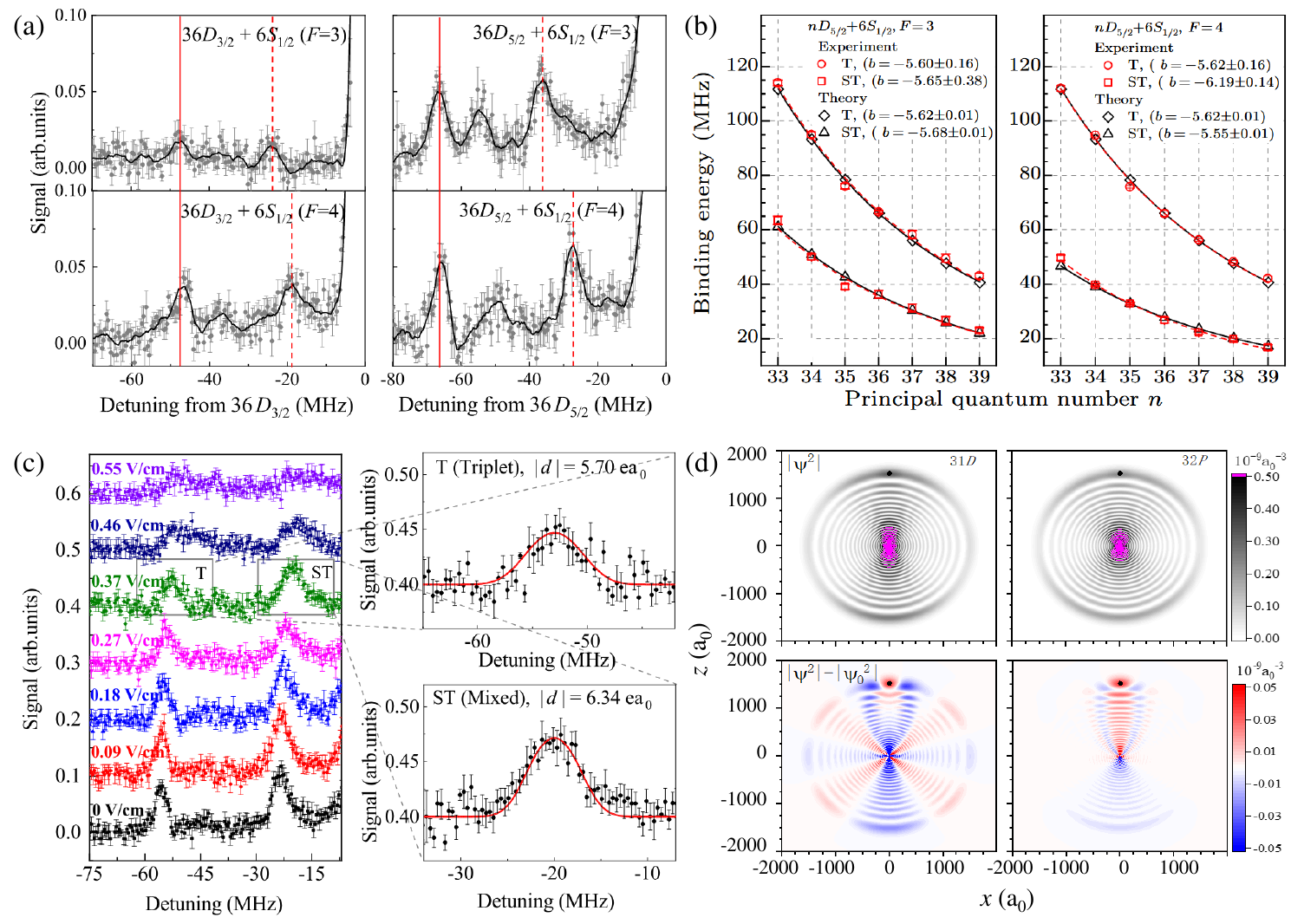}
\end{center}
\caption{ (a) Measured molecular spectra of $36D_J-6S_{1/2}$ molecules for $J=3/2$ (left) and $J=5/2$ (right), for $F=3$ (top) and $F=4$ (bottom). The vertical dashed and solid lines mark the signals of the $^{S,T}\Sigma~(\nu=0)$ and $^T\Sigma~(\nu=0)$ vibrational states, respectively~\cite{baiPRR2020}. (b) Measured (red) and calculated (black) binding energies of the ground vibrational states of $nD_{5/2}-6S_{1/2}$ ground-Rydberg molecules for $F = 3$ (left) and $F = 4$ (right)~\cite{baicpl2020}. (c) Spectra of $37D_{5/2} - 6S_{1/2}~(F=4)$ ground-Rydberg molecules with the indicated electric fields. The red solid lines show the fitting model of dipole moments, yielding $|d| = 5.70~ea_0$ for $^T\Sigma~(\nu=0)$ and $6.34~ea_0$ for $^{S,T}\Sigma~(\nu=0)$, respectively~\cite{baiPRR2020}. (d) Calculations of electronic densities for Cs $31D_{5/2}-6S_{1/2}$ (left) and $32P_{3/2}-6S_{1/2}$ (right) with a perturber at $\approx 1500~a_0$. Top: wave-function densities; bottom: molecular–atomic differences. Constructive ($P$-type) and destructive ($D$-type) trilobite interference yield dipole moments of $7~ea_0$ and $−5~ea_0$, respectively~\cite{baiPRR2020}. (Reproduced with permission and adapted from Ref.~\cite{baiPRR2020} and ~\cite{baicpl2020}, licensed under a Creative Commons Attribution 4.0 International license.)
}\label{Fig6}
\end{figure}

Figure~\ref{Fig6}(a) presents the measured ceium $(6S-nD_{J})$ ground-Rydberg molecular spectra for $n = 36$ employing the two-photon photoassociation technique~\cite{baiPRR2020}. Considering the fine structure interaction of Rydberg atoms and hyperfine structure interactions of ground-state atoms, photo-association spectra of $D$-state ground-Rydberg molecules for all combinations of $J$ and $F$ are obtained. The laser detunings are relative to the atomic resonances, and the signal strengths are displayed on identical scales. Vertical solid lines mark the molecular signals of $^T\Sigma (\nu=0)$ ground vibrational states, while vertical dashed lines mark the signals of the $^{S,T}\Sigma (\nu=0)$ ground vibrational states. The signal strengths of the $J = 5/2$ spectra are higher than those of the $J = 3/2$ ones, due to the higher
excitation probability of the $nD_{5/2}$ atoms. From Figure~\ref{Fig6}(a) we can see that the deep potentials mostly arise from triplet $s$-wave scattering ($^T\Sigma$ )
and do not depend on $F$ quantum number, whereas the shallow potentials mostly arise from $s$-wave scattering of mixed $^{S,T}\Sigma$ states and depend on $F$ quantum number, and binding energy for $^{S,T}\Sigma,F = 3$ is deeper than that for $^{S,T}\Sigma,F = 4$. 

In Figure~\ref{Fig6}(b), we present the binding energy of Rydberg molecules for all combinations of $J$ and $F$ as a function of principal quantum number for $n$=33 to 40~\cite{baicpl2020}. By fitting the experimental data, the scaling law of the binding energy can be obtained, as shown in the blank area of the figure. By comparing the experimental observations and calculations, the zero-energy singlet and triplet $s$-wave scattering lengths can also be extracted. 

As mentioned above, the ground-Rydberg molecule has a permanent dipole, and the analysis of the spectral line broadening in the applied external electric field yields a permanent dipole moment, as shown in Figure~\ref{Fig6}(c)~\cite{baiPRR2020}. By fitting and analyzing the broadening of molecular signal spectral lines, the averaged dipole-moment were $(4.79 \pm 0.78)~ea_0$ for $^T\Sigma$ and $(5.49 \pm 1.03)~ea_0$ for $^{S,T}\Sigma$ of $37D_{5/2} + 6S_{1/2}(F = 4)$ Rydberg molecules. However, in contrast to the previously reported positive dipole moment, the calculated dipole moment for $D-$type ground-Rydberg molecule are negative. This is due to the perturbation of the ground state atoms on the nearby Rydberg electron density, which leads to the destructive interference of the nearby wave functions, resulting in a negative permanent electric dipole moment, as shown in Figure~\ref{Fig6}(d), where the calculated trilobite orbital of the $D$-type molecule predominantly shows destructive interference with the $D$ orbital, causing a negative dipole moment of approximately 5~$ea_0$~\cite{baiPRR2020,jiao2023}. Besides, the decay rates of cesium $nD_{5/2}+ 6S_{1/2}$ ground-Rydberg molecules are much faster than that of the corresponding parent Rydberg atoms~\cite{baijxjcp2023}, which is different from rubidium~\cite{Butscher2011} and strontium~\cite{Camargo2016} Rydberg-ground molecules. This is attributed to the four dissociation channels of collision, blackbody radiation, tunneling, and coupling of neighboring Rydberg states.
 
\begin{figure}[htbp]
\begin{center}
\includegraphics[width=0.4\textwidth]{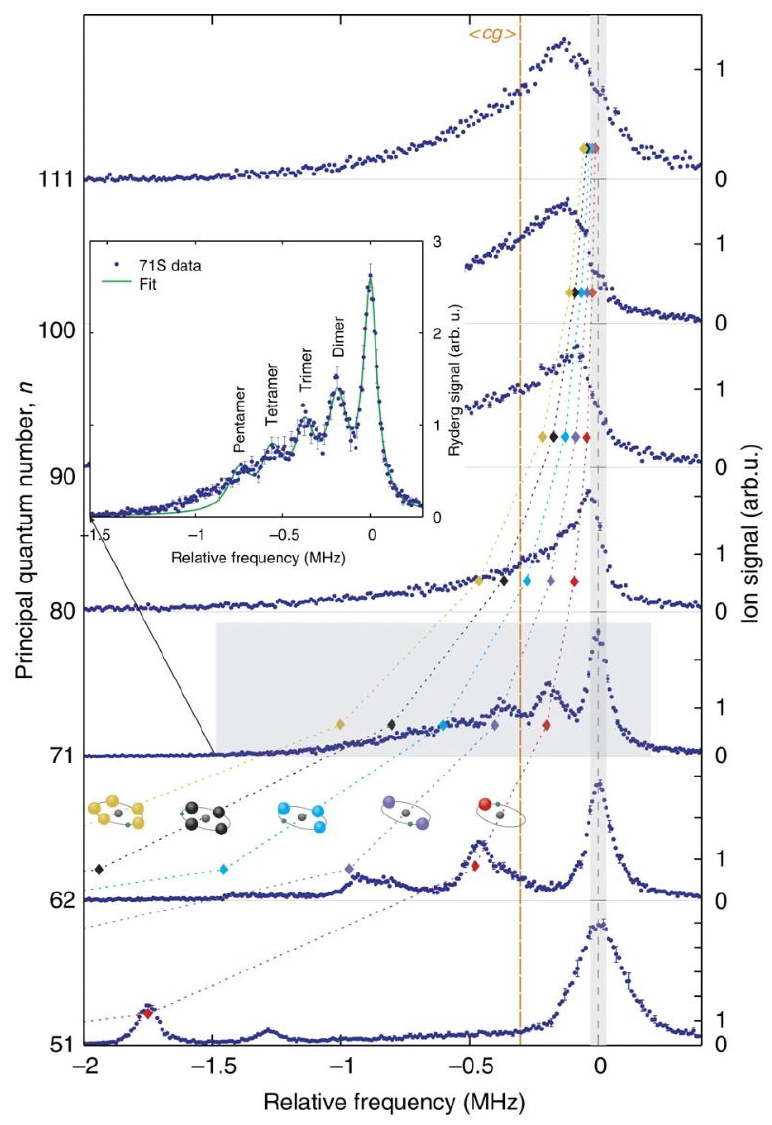}
\end{center}
\caption{ $5S_{1/2}\to nS_{1/2}$ excitation spectra showing dimer and polyatomic Rydberg molecular lines emerging with increasing $n$. Frequencies are referenced to the atomic resonance for $n\leq71$ (dashed line); the grey band indicates the laser bandwidth. Spectra are aligned by their centres of gravity, assuming constant density shifts. Molecular resonances up to three ($62S$) and four ($71S$) bound ground-state atoms are resolved and follow a power-law scaling of binding energies. The inset shows the $71S$ spectrum with a multi-Lorentzian fit. The spectra are averaged over 20 measurements with standard-deviation error bars~\cite{gajnc2014}. (Reproduced with permission and adapted from Ref.~\cite{gajnc2014}, licensed under a Creative Commons Attribution 4.0 International license.) }\label{Fig7}
\end{figure}

Apart from diatomic molecules, the low-energy scattering interaction can be extended to bind multiple ground-state atoms~\cite{Liu2006,Liu2009,Bendkowsky2010,gajnc2014,Eiles2016,Fey2016,Luukko2017,fey2019a,Chien2024}, forming the polyatomic Rydberg molecule.  Figure~\ref{Fig7} shows the photo-association spectra of multi-atomic ground-Rydberg molecules, investigated by the research group of the University of Stuttgart~\cite{gajnc2014}. With increasing $n$, transitions associated with dimer and polyatomic Rydberg molecules become clearly visible. Multi-atomic molecules, including dimers (red), trimers (violet), tetramers (blue), pentamers (black), and hexamers (yellow) of ground-Rydberg molecules, are marked by rhombuses. The investigation showed that the binding energy of multi-atomic polymers is in a linear proportional relationship with the number of ground-state atoms, that is, the binding energy is an integer multiple of the binding energy of double-atomic Rydberg-ground-state molecules. This discovery makes Rydberg molecules an ideal system for studying quantum correlations from few-body to many-body systems, effective many-body potentials, and many-body binding mechanisms.

Further investigations have also been carried out on ground-Rydberg molecules formed from high-angular-momentum Rydberg atoms. The vibrational spectrum of the rubidium $22F$ ground–Rydberg molecule reveals vibrational levels up to 
$\nu = 6$, in good agreement with the calculated potential energy curves and vibrational wave functions for both single-triplet and mixed single-triplet potential wells~\cite{Althon2023,Exner2025}. These molecules exhibit permanent electric dipole moments on the order of kilo-Debye and, moreover, possess lifetimes longer than that of the parent Rydberg atom.

Recently, alkaline earth metal atoms with two valence electrons, such as Sr~\cite{Whalen2020,Lu2022} and Yb~\cite{Legrand2025} have been ideal platforms for investigating Rydberg atoms and molecules. Owing to their two-valence-electron configuration, alkaline-earth Rydberg atoms can be described by a single excited Rydberg electron coupled to an inner electron, which introduces additional internal degrees of freedom while maintaining ultralong-range molecular bindings. This structure enables optical access to the inner electron, facilitating high-precision spectroscopy and coherent control. In addition, their relatively small quantum defects, weak spin–orbit coupling, and narrow-line optical transitions lead to more regular Rydberg spectra, allowing accurate theoretical modeling and high-resolution photo-association spectroscopy, making alkaline-earth Rydberg molecules promising candidates for precision measurements and quantum applications.

Ground-Rydberg molecules have been widely investigated with different atoms, such as Rb and Cs, different angular momentum states, such as $nS$, $nP$, $nD$ and $nF$ states, as well as different alkaline-earth metal atoms, such as Sr and Yb atoms, and other compounds, such as NO~\cite{Gonz2021,Rayment2021,Deller2020}. Significant progress has been made in the study of ground-Rydberg molecules both in experiments~\cite{Kleinbach2017,Engel2019,Peper2020} and in theoretical simulations~\cite{Luukko2017,Schmid2018,Giannakeas2020,Hummel2021,Hummel2023,Srikumar2023,Eiles2024,Durst2025} by advanced calculation method and quantum technology like optical tweezers arrays~\cite{Mellado2024}, revealing their remarkable properties. Key research advancements include a deeper understanding of intrinsic molecular properties—such as lifetime, binding energy, and dipole moment—as well as more quantitative models of bound states. Further work has enabled the investigation of the low-energy electron-atom collisions, the external control of molecular via electric and magnetic fields~\cite{Hummel2019}, and the probing of spatial correlations within ultracold atomic gases.

\section{Rydberg macrodimer}\label{sec3}

Rydberg-Rydberg molecule, also called Rydberg macrodimer, formed with two Rydberg atoms~\cite{Boisseau2002}. The binding mechanism of this type of molecule involves multiple electrostatic interactions between two or more Rydberg atoms. Since the interaction between Rydberg atoms occurs at distances extending beyond the Rydberg orbital for identical principal quantum numbers, the Rydberg macrodimer exhibits large size and its bond lengths scale approximately as $n^{2.5}$ and exceed the LeRoy radius $\sim 4n^2$, and can easily exceeds $1~\mu$m. Thereby Rydberg marodimer is the largest known diatomic molecule in existence~\cite{Shaffer2018}.

\subsection{Binding mechanism}\label{sec3.1}

\begin{figure}[htbp]
\begin{center}
\includegraphics[width=0.3\textwidth]{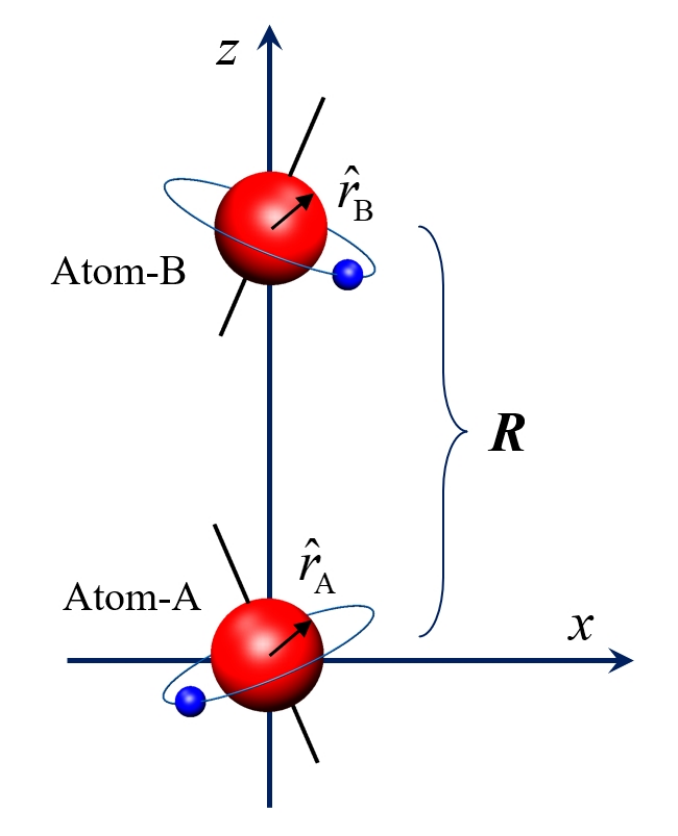}
\end{center}
\caption{ Two-atom interaction model. The Rydberg atoms~A (located at the origin) and~B are separated by a distance {\textbf{R}} along the $z$-axis. $\hat{r}_A$ and $\hat{r}_B$ represent the relative positions of the Rydberg electrons in atoms~A and~B, respectively. }\label{Fig8}
\end{figure}

Rydberg atoms have strong long-range interactions between atoms, scaling as $n^{11}$. Boisseau et al. investigated the interaction of $P$-state atoms and found that the long-range interaction of $P$-state Rydberg atoms could form a bound potential well, which firstly predicted the existence of long-range Rydberg molecules~\cite{Boisseau2002}. To better understand the binding mechanism of the Rydberg macrodimer molecules, we calculate the interaction of a Rydberg-atom pair. We consider the theoretical model as shown in Figure~\ref{Fig8}. Considering two atoms $A$ and $B$ separated by a distance $\textbf{R}$, both of which are excited to Rydberg states. For simplicity, we assume that atom $A$ is located at the origin of the coordinate system, with the quantization axis chosen along the $z$ direction, while atom $B$ is positioned on the $z$ axis. The vectors $\mathbf{r}_A$ and $\mathbf{r}_B$ denote the relative positions of the Rydberg electrons with respect to their respective ionic cores. Within the Born-Oppenheimer approximation~\cite{Born1927Zur}, the total Hamiltonian of the two Rydberg atoms can be written as: 
\begin{equation}
\hat{H} = \hat{H}_A + \hat{H}_B + \hat{V}_{\mathrm{int}},\label{eq4}
\end{equation}
where $\hat{H}_{A(B)}$ denotes the Hamiltonian of individual atoms, while $\hat{V}_{\text{int}}$ represents the Hamiltonian of the interaction between two atoms.

Multipolar interactions constitute the primary intermolecular forces and play a pivotal role in the formation of Rydberg macrodimers. To accurately determine the magnitude of these interactions, higher-order multipole terms are typically incorporated in the calculations. Assuming that the distance between atoms is greater than Le Roy's radius~\cite{LeRoy1974Long}, ensure that the electron wave functions do not overlap~\cite{Jackson1998Classical}. The effects of exchange and charge overlap interactions were ignored ~\cite{Casimir1948Influence}. Thus, the Rydberg electrons can be regarded as distinguishable particles. The interaction term $\hat{V}_{\text{int}}$ can be expanded in multipole series and expressed in atomic units as~\cite{Schwettmann2006Cold,Deiglmayr2014Observation,Deiglmayr2016Long,Han2018Cs,Han2019Adiabatic}:
\begin{equation}
\hat{V}_{int} = \sum_{q=2}^{q_{\max}} \frac{1}{R^{q+1}} \sum_{L_A=1,L_B=q-L_A}^{q  _{\max}-1} \sum_{Q=-L_<}^{L_<} f_{ABQ} \hat{Q}_A \hat{Q}_B,\label{eq5}
\end{equation}
where $L_{A(B)}$ denotes the order or rank of the multipole interaction, $L_<$ represents the smaller value between $L_A$ and $L_B$, $q=L_A+L_B$ is the total interaction order. $\hat{Q}_{A(B)}$ and $f_{ABQ}$ denote the interatomic electrostatic interaction operator and factor, defined as:
\begin{equation}
\hat{Q}_A = \sqrt{\frac{4\pi}{2L_A + 1}} \hat{r}_{A}^{L_A} Y_{L_A}^{\Omega}(\hat{r}_A),\label{eq6}
\end{equation}

\begin{equation}
\hat{Q}_B = \sqrt{\frac{4\pi}{2L_B + 1}} \hat{r}_{B}^{L_B} Y_{L_B}^{\Omega}(\hat{r}_B),\label{eq7}
\end{equation}

\begin{equation}
f_{ABQ} = \frac{(-1)^{L_B} (L_A + L_B)!}{\sqrt{(L_A + \Omega)!(L_A - \Omega)!(L_B + \Omega)!(L_B - \Omega)!}},\label{eq8}
\end{equation}
where $\hat{r}$ denotes the radial matrix element, $Y$ represents the spherical harmonic function, $\Omega$ signifies the counting index. By numerically solving the Hamiltonian of the Rydberg atom pair on a dense grid of the internuclear separation, $R$, we can obtain the adiabatic molecular potential energy curve of the Rydberg atom pair-state.

\subsection{Potential energy curves (PECs)}\label{sec3.2}

Considering the angular symmetry, the total angular momentum $M = m_{jA} + m_{jB}$ is conserved, where $m_j$ is the magnetic quantum number of the electrons in the Rydberg atoms. The Hamiltonian represented by equation~(\ref{eq4}) is diagonalized, and the energy of the Rydberg atomic pair-state is numerically calculated when the atomic nucleus distance is $\textbf{R}$. This yields the adiabatic potential energy curve of the Rydberg atomic pair.
  
\begin{figure}[htbp]
\begin{center}
\includegraphics[width=0.7\textwidth]{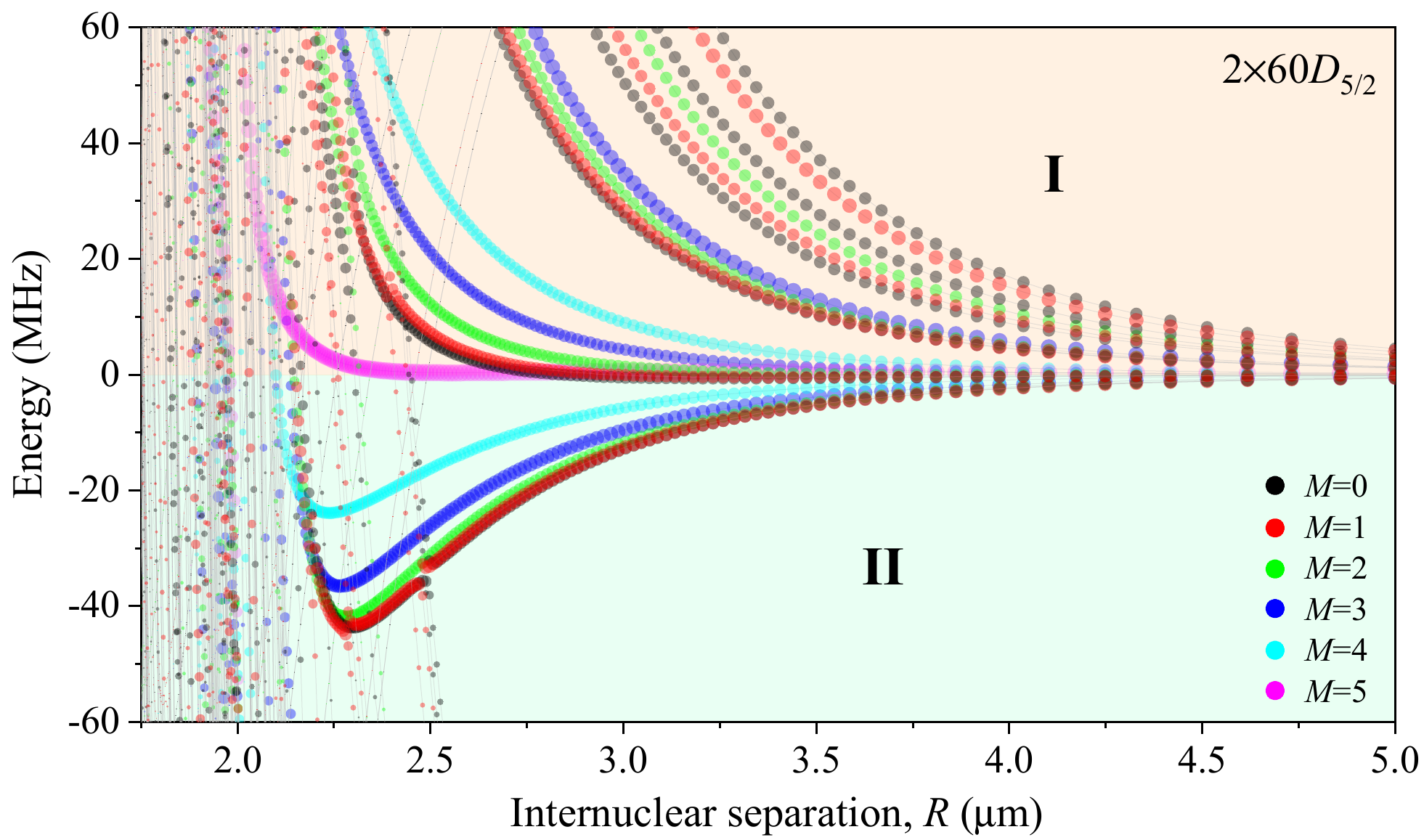}
\end{center}
\caption{Adiabatic potential energy curves of $60D_{5/2}$ Rydberg-atom pair for indicated $M$ values, symbol areas are proportional to laser excitation rates, and symbol colors correspond to different $M$ values. }\label{Fig9}
\end{figure}

To calculate the adiabatic potential energy curve, truncation parameters such as the interaction order and basis set and single-atom orbital angular momentum space are carefully choose. The interaction order, $q_{\text{max}}=6$, encompasses dipole-dipole, dipole-quadrupole, quadrupole-quadrupole, and dipole-octupole interactions. The parameter range of the quantum number satisfies: $(\text{Int}(n_{eff})-\Delta) < n_{eff} < (\text{Int}(n_{eff})+\Delta+1)$, where $\text{Int}(n_{eff})$ represents the integer operation of the effective principal quantum number $n_{eff}$, and $\Delta$ is the range of the principal quantum numbers included in the calculation. Here, $\Delta=3.1$. The maximum values of the orbital angular momentum quantum number and magnetic quantum number for a single atom are $l_{\text{max}} = 5$ and $m_j = 5$, respectively. The maximum energy detuning of the state atom energy is 30~GHz~\cite{BjxPhD2024}.
  
In Figure~\ref{Fig9}, as an example, we present the numerically computed adiabatic molecular potential energy curves for the cesium atom $2\times60D_{5/2}$ Rydberg pair state with different $M$ values~\cite{Han2019Adiabatic,baijx2018b}, 
the zero energy is defined to be the $60D_{5/2}60D_{5/2}$ pair asymptote. The PECs can be divided into two regions I and II based on energy. In region~I, the upward-bending adiabatic potential energy curves of the Rydberg atom pair exhibit repulsive potentials, which do not contribute to the formation of Rydberg molecules. However, in regionII, the downward-bending potential energy curves display attractive potentials, featuring potential minima that can form bound potential wells, which are larger than several dozen MHz, thereby binding the adjacent Rydberg atom to form Rydberg molecules. Besides, as the distance between two atoms decreases, the numerous gray potential energy curves within the internal region about $R<2.25~\mu$m represent the extra repulsive interactions between other Rydberg atoms. The area of the colored symbols in Figure~\ref{Fig9} is proportional to the excitation probability of the Rydberg macrodimers~\cite{Han2019Adiabatic}. 

\begin{figure}[htbp]
\begin{center}
\includegraphics[width=0.8\textwidth]{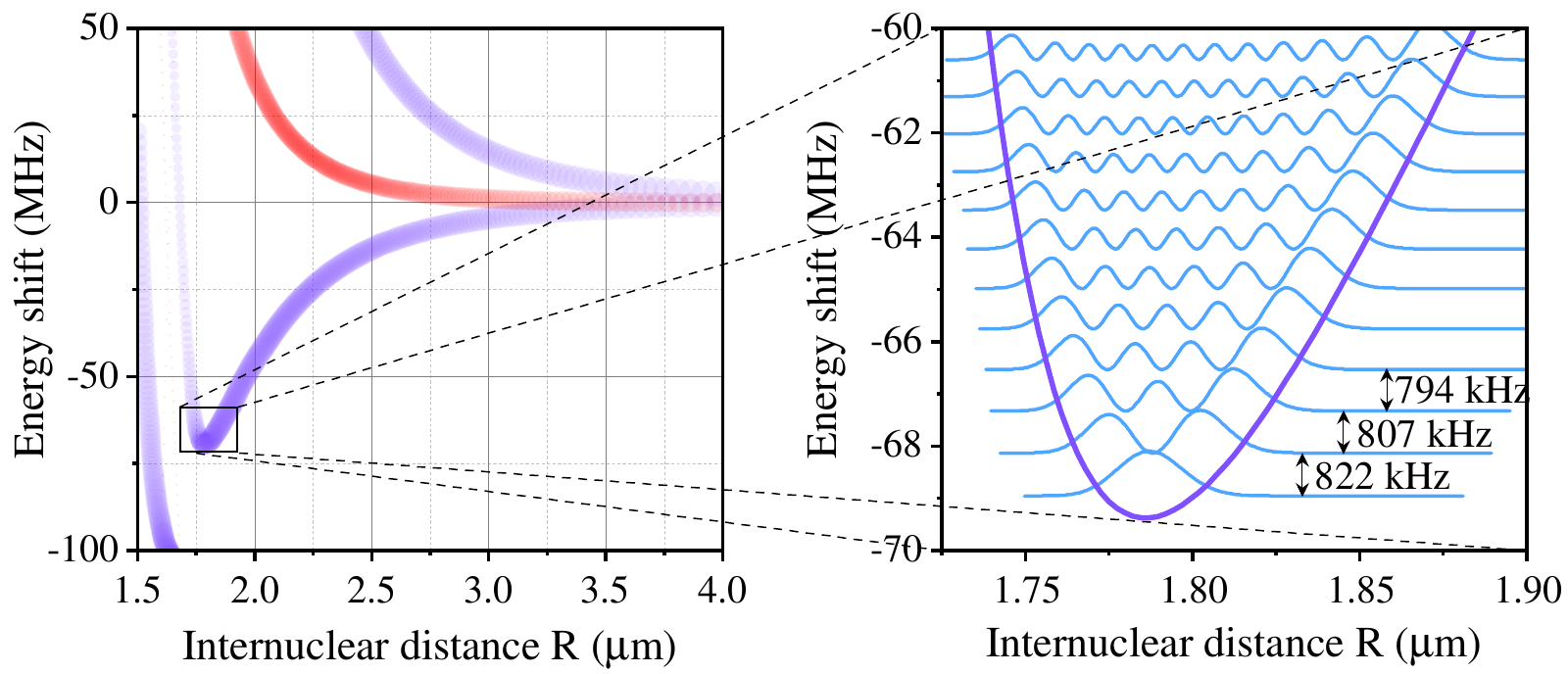}
\end{center}
\caption{Adiabatic potential energy curve of $(60D_{5/2})_2$ Rydberg-atom pair for indicated $M=3$ and for the even parity (red) and odd (purple), symbol areas are proportional to laser excitation rates. The right panel shows an enlarged view of the molecular potential well within the black box area on the left, along with its vibration wave function~\cite{Han2019Adiabatic}. }\label{Fig10}
\end{figure}

The potential wells in the Rydberg molecular potential energy curves are typically tens of MHz deep and can support several bound vibrational states. This feature makes Rydberg molecules particularly attractive for laser-spectroscopic investigations and time-dependent wave-packet studies. Based on the Rydberg molecular PECs and the molecular Hamiltonian theory~\cite{yang2008,yang2010,yang2012}, we can compute the vibrational energy levels and vibrational wave functions of the molecule. As an example, the potential energy curves of $(60D_{5/2})_2,~M=3$ are shown as Figure~\ref{Fig10}~\cite{Han2019Adiabatic}. We can see that, among the four potential energy curves with a relatively high excitation probability, only one shows an attractive potential, and there is a potential minimum of approximately $-69$~MHz, which represents a molecular binding potential well, capable of binding adjacent Rydberg atoms to form Rydberg molecules. The other three potential energy curves represent repulsive potentials and do not form binding potential wells. And the red and purple colors represent even parity and odd parity, respectively. 

The black box marks the molecular potential well. The right panel shows the enlargement and calculated vibrational energy levels and vibrational wave functions. It can be seen that the giant Rydberg molecule has many vibrational energy levels. The 10~MHz potential well contains 12 vibrational energy levels. The intervals between adjacent vibrational energy levels are less than 1~MHz, and the interval between the vibration ground state and the first vibrational energy level is 822~kHz. For non-harmonic potential wells, the intervals between adjacent vibrational energy levels decrease as the vibration quantum number increases. Finally, for diatomic molecules, the additional degrees of freedom, such as vibration and rotation in these molecules, increase the complexity of the molecular state, resulting in different symmetries of the molecular quantum state, mainly including rotational symmetry, inversion symmetry, reflection symmetry, and the combination of rotation and reflection~\cite{Samboy2011,Samboy2017,Weber2017Calculation,hollerithPhD2022}. 

The coupling of angular momenta in molecules is highly intricate, as all angular momenta and rotational angular momenta of individual atoms contribute to the molecular angular momentum. Depending on the coupling strength and molecular binding energy, different projections of angular momentum can be assumed to be conserved. This idealized scenario is referred to as Hund's cases~\cite{Hund1933,Brown2003}. As shown in Figure~\ref{Fig9} above, the Hund's case~$\mathbb{C}$ including even and odd parity is adopted to characterize the cesium Rydberg macrodimers. Rydberg-Rydberg molecules are weakly bound molecules, and their binding energy is lower than that of $LS$ coupling, so the coupling of rotational angular momentum can be ignored. For deeply bound molecules, other projections such as $\Lambda=m_{L1}+m_{L2}$ and $\Sigma=m_{S1}+m_{S2}$ can also be assumed to be conserved, which is applicable to other Hund's cases~$\mathbb{A}$ and~$\mathbb{B}$~\cite{hollerithPhD2022}.

\subsection{Experimental Realization}\label{sec3.3}

\begin{figure}[htbp]
\begin{center}
\includegraphics[width=0.7\textwidth]{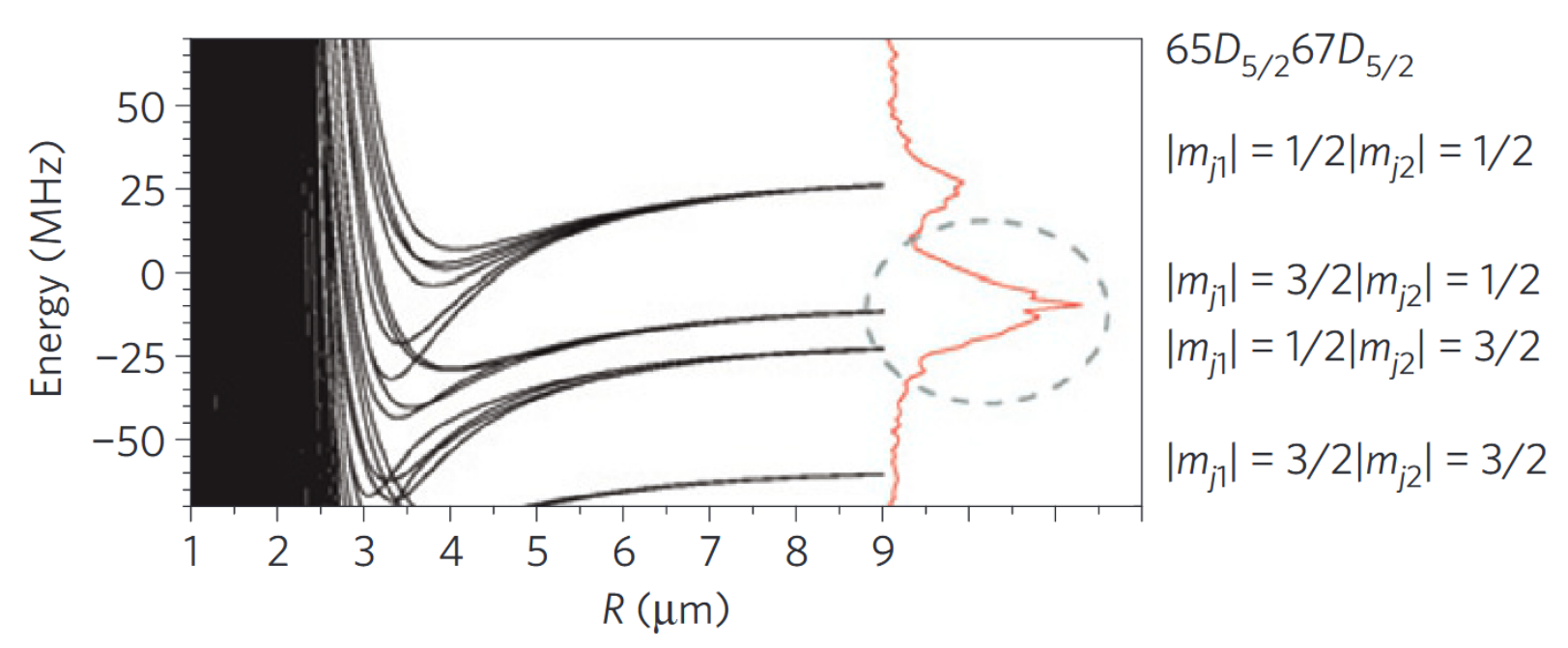}
\end{center}
\caption{ Integrated atomic ions yield spectra as a function of laser detuning (orange curve), overlaid with calculated pair potentials for the $65D-67D$ pair at an electric field of $190~\text{mV}~\text{cm}^{-1}$. All fine-structure components and molecular states $M=m_{j1}+m_{j2}$ are included, where $M$ denotes the projection of the total angular momentum onto the internuclear axis $\textbf{R}$. The dashed circle marks one of the molecular features analyzed in the experiment~\cite{overstreet2009}. (Reproduced with permission and adapted from Ref.~\cite{overstreet2009}, licensed under a Creative Commons Attribution 4.0 International license.)}\label{Fig11}
\end{figure}

The first experimental evidence of Rydberg macrodimers of $nD_{5/2}(n+2)D_{5/2}$ was observed in a cesium MOT in 2009 by studying the field ionization time-of-flight spectra of Rydberg atoms~\cite{overstreet2009}. Figure~\ref{Fig11} presents the spectrum and calculated potentials for the $65D + 67D$ pair state within an electric field of $190~\text{mV}~\text{cm}^{−1}$. The pair interaction potentials exhibit pronounced wells at internuclear separations of $R \approx 3\sim 9~\mu$m, which can support hundreds of bound vibrational states with maximum level spacings on the order of 100~kHz. The molecular lifetimes are primarily limited by radiative and blackbody-induced decay of the constituent Rydberg atoms. Since the decay of either atom leads to molecular dissociation, the molecular lifetimes are approximately half those of the corresponding atomic states. At 300~K, the lifetimes of the $65D+67D$ molecular pairs are approximately $46~\mu$s. Subsequently, Saßmannshausen et al. directly observed the spectral signals of Rydberg macrodimers of $nP(n+1)S$ and $nSn'F$ using the photo-association spectra of Rydberg atoms~\cite{Deiglmayr2014Observation,Sassmannshausen2016}, particularly at laser detunings close to the bound potential well position. These experiments once again confirmed the existence of the Rydberg macrodimers.

\begin{figure}[htbp]
\begin{center}
\includegraphics[width=0.85\textwidth]{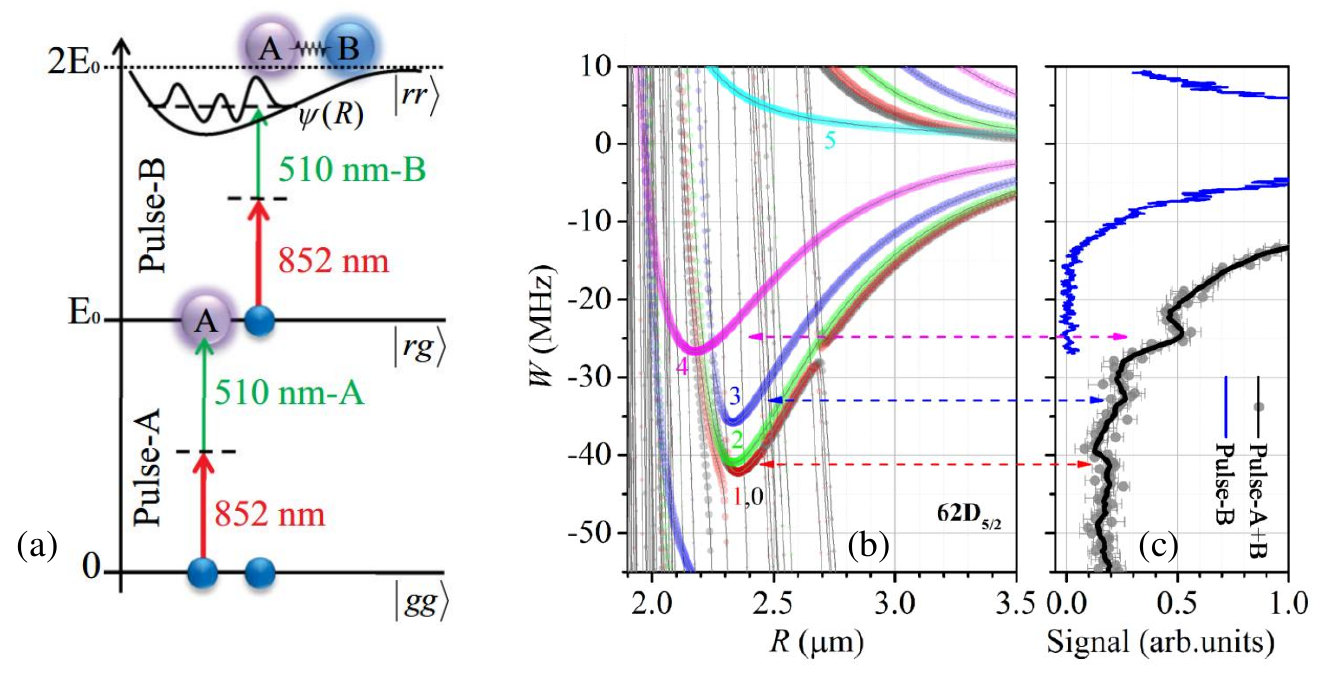}
\end{center}
\caption{ (a) Level diagram and sketch of a vibrational wave function for two-color double-resonant excitation of $(62D_J)_2$ Rydberg-atom macrodimers. The pulse pair A resonantly excites the seed Rydberg atoms (atom A). To study Rydberg-atom macrodimers, the frequency of the 510~nm component of the pulse pair B is scanned relative to the atomic resonance. (b) Calculated adiabatic potentials (gray) for cesium Rydberg macrodimers $(62D_{5/2})_2$ at different $M$. Symbol sizes indicate excitation rates averaged over random molecular alignment; colors denote $|M|$. (c) Single-color spectrum (pulse B, blue) and two-color macrodimer spectrum (pulses A+B, symbols; black line: smoothed average). The single-color peak marks the $(62D_{5/2})_2$ asymptote ($W=0$). Dashed lines indicate the minima of the two molecular potentials~\cite{Han2018Cs}. (Reproduced with permission and adapted from Ref.~\cite{Han2018Cs}, licensed under a Creative Commons Attribution 4.0 International license.)}\label{Fig12}
\end{figure}

Due to the weak binding of the Rydberg macrodimer molecules, the photoassociation rate of the molecules in the experiment was relatively low. Han et al. proposed a two-color two-photon photo-association approach to increase the photo-association rate of Rydberg macrodimer molecules and successfully synthesized Rydberg macrodimers of cesium atoms with orbital angular momentum $L=2$~\cite{Han2018Cs}. The left side of Figure~\ref{Fig12}(a) shows the level diagram of two-step, two-color photo-association. The seed Rydberg atom~A is resonantly excited from the ground state to the Rydberg state with a laser pulse~A. The second laser pulse~B is detuned relative to that of pulse~A, and excited Rydberg atoms close to the seed atom~A at a distance where metastable $(62D_{5/2})_2$ macrodimers exist. Owing to the doubly resonant character of this two-color photo-association scheme, the excitation rate is greatly enhanced in comparison with that of single-color photo-association. Figure~\ref{Fig12} (b, c) presents the experimentally measured photo-association spectra alongside the theoretically calculated adiabatic molecular potential energy curves, with distinct molecular spectral peaks marked by colored vertical dashed lines, three molecular peaks correspond to $M$=4 (pink), 3 (blue) and 0 (1 and 2) (red), respectively. The potential well depths for $M$=0, 1, and 2 are nearly the same and cannot be distinguished in the experiment. The molecular state of $M=5$ shows an upward repulsive interaction, which makes it impossible to form a potential well and trap the Rydberg macrodimers. The measured photo-association spectra of Rydberg molecules show a agreement with the theory. By measuring the molecular ion signal versus the waiting time between the laser pulses and ionization field, Rydberg macrodimers expressed a fast decay on time scales of several microseconds. In addition, building upon the existing theory of electric multipole interactions, they conducted a comprehensive theoretical analysis of molecular adiabatic potential energy curves by incorporating higher-order interactions and basis vector dimensions. This analysis yielded key spectroscopic characteristics, including the depth of the binding well, equilibrium internuclear distance, and vibrational wave functions. 

\begin{figure}[htbp]
\begin{center}
\includegraphics[width=0.8\textwidth]{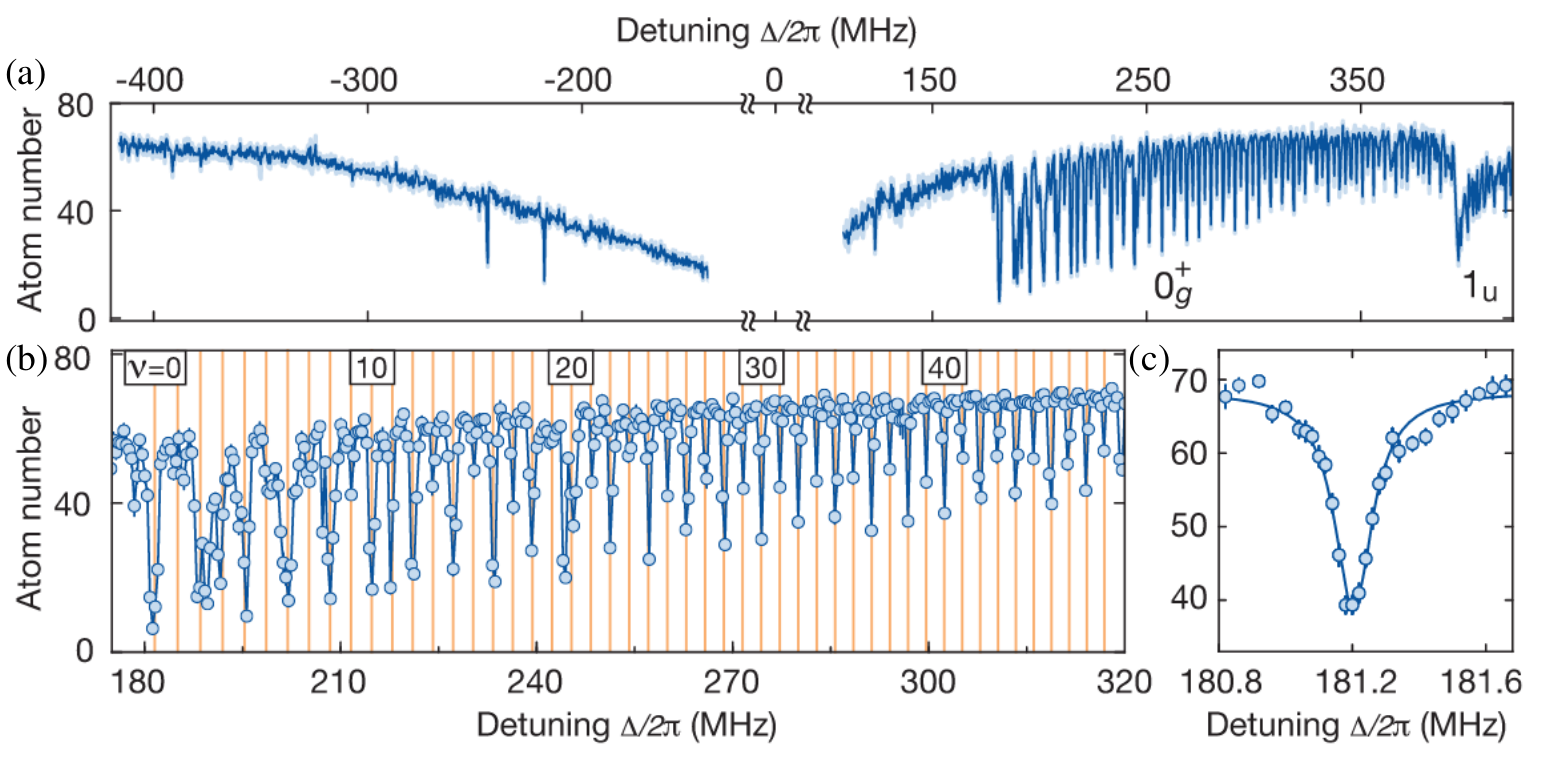}
\end{center}
\caption{ (a) Wide-range spectrum showing no bound states on the red-detuned side, while regularly spaced loss features on the blue-detuned side indicate vibrational resonances of Rydberg macrodimers; an additional series from a $1_u$ potential appears at larger detunings. (b) Zoom-in between 180 and 320~MHz reveals 3~MHz vibrational spacing, which decreases slightly for higher states and is in good agreement with the theory (orange lines). (c) Spectrum of the lowest vibrational level. Error bars denote the SEM~\cite{hollerith2019science}. (Reproduced with permission and adapted from Ref.~\cite{hollerith2019science}, licensed under a Creative Commons Attribution 4.0 International license.)}\label{Fig13}
\end{figure}

In recent years, the Bloch research group in Germany employed a quantum gas microscope to investigate Rydberg macrodimers with unprecedented microscopic resolution~\cite{hollerith2019science,Hollerith2022Realizing,Hollerith2023Rydberg,hollerith2024c}. Starting from ground-state $^{87}\text{Rb}$ atoms deterministically arranged in a two-dimensional optical lattice, they photoassociated pairs of atoms into Rydberg macrodimers via two-photon excitation~\cite{hollerith2019science}. Figure~\ref{Fig13}(a) presents the high-resolution two-photon spectroscopy performed at frequencies detuned from the atomic Rydberg resonance. Over a wide frequency range, no signatures of molecular bound states are observed on the interaction-broadened red-detuned side of the Rydberg resonance, but the two isolated spectral features appearing in this region can be attributed to lattice-induced Raman resonances. In contrast, on the blue-detuned side, a series of pronounced and regularly spaced loss features emerges, which are identified as the vibrational resonances of Rydberg macrodimers. A closer inspection of the spectral window between 180~and 320~MHz reveals more than fifty vibrational bound states of the macrodimers spacing of approximately 3~MHz, with a slight decrease in spacing for higher-lying vibrational states. The experimentally observed resonance positions are in very good agreement with ab initio theoretical predictions (orange lines). The small image on the right shows the high-resolution spectroscopy of the lowest vibrational level. In addition, by leveraging single-atom sensitivity and site-resolved fluorescence imaging, they directly identified macrodimer formation as the correlated loss of atom pairs separated by a characteristic bond length, and further demonstrated control over the spatial orientation of the molecules through the vibrational parity and laser polarization. They further demonstrated possible schemes for multi-body physics, Rydberg decoration, and entanglement by coupling the molecular potentials between Rydberg atomic pairs through non-resonant laser coupling and designing interatomic interactions. The decoherence effect of weakly bound Rydberg atoms and Rydberg macrotrimer molecules were further investigated. 

\begin{figure}[htbp]
\begin{center}
\includegraphics[width=0.6\textwidth]{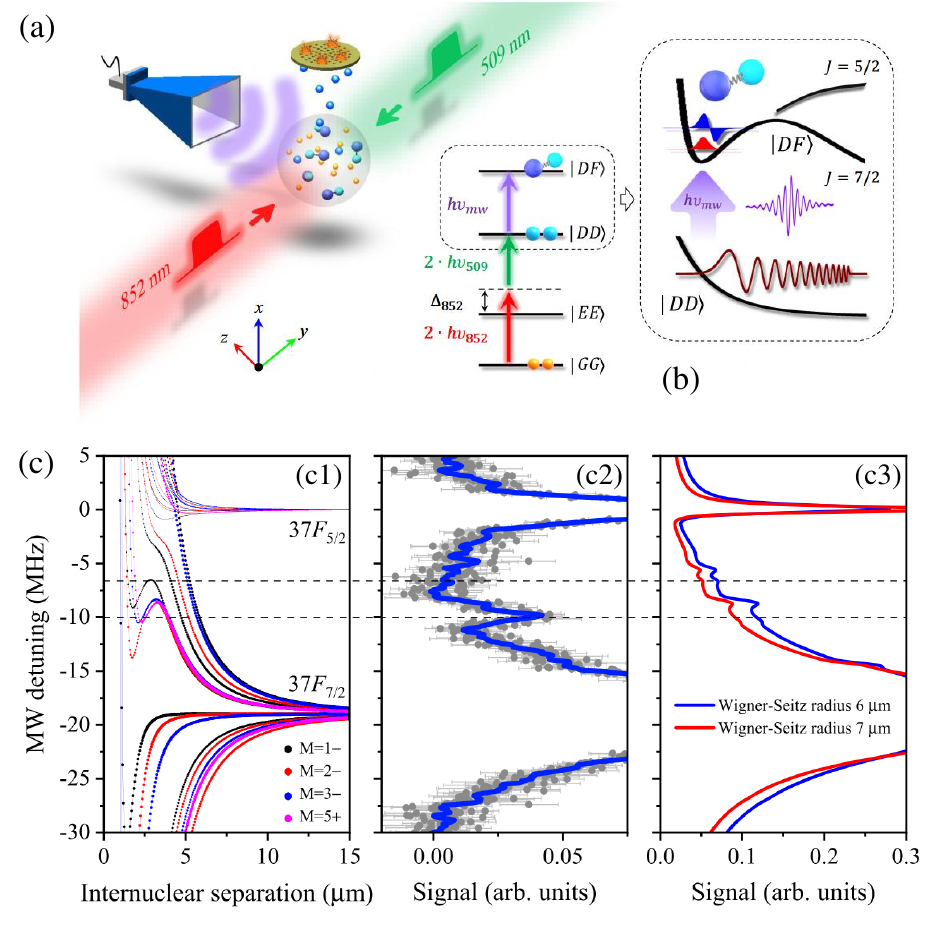}
\end{center}
\caption{ (a) Sketch of the experimental setup. Rydberg macrodimers are produced by laser (852~nm and 509~nm) and microwave photoassociation and detected by a microchannel plate detector (MCP). (b) Level diagram. A Rydberg pair, $|DD\rangle$, is laser-excited from $|GG\rangle$ by 852 and 509~nm lasers (852~nm laser detuning $\Delta_{852}=2\pi\times 360$~MHz). A scanned microwave field excites $nF_J$ atoms or photo-associates $|DF\rangle= (n+2)D_{5/2}nF_J$ fine-structure (FS)-mixed macrodimers. (c) Theoretical calculation of the potential energy curve and the measurement and simulation of microwave photoassociation spectra. (c1) Calculated potential energy curves vs internuclear separation $R$ for $39D_{5/2}37F_J$ pairs with the indicated molecular state $M$ and parity ( $−/+$ means odd/even). Symbol areas are proportional to  calculated microwave transition rates from the launch state $2\times39D_{5/2}$. (c2) Experimental microwave spectrum, averaged over 50 measurements. The signal at microwave detuning $-10$~MHz reflects FS-mixed $39D_{5/2}37F_J$ macrodimers. (c3) Simulated microwave spectra for two Wigner-Seitz radii~\cite{baiprr2024}. (Reproduced with permission and adapted from Ref.~\cite{baiprr2024}, licensed under a Creative Commons Attribution 4.0 International license.)
}\label{Fig14}
\end{figure}

In the early stage, one mainly focused on the Rydberg macrodimers for the Rydberg states with lower orbital angular momentum, which are prepared by the laser photo-association method. By combining optical excitation and microwave coupling in a cooperative manner, Bai et al. reported a new type of Rydberg macrodimers, the fine-structure-mixed (FS-mixed) long range $(n+2)D_{5/2}-nF_J$ Rydberg macro-dimer molecules~\cite{baiprr2024}. Figure~\ref{Fig14}(a) and (b) present the experimental setup and energy level diagram for the preparation of Rydberg macrodimers through microwave photo-association. Two excitation laser pulses and a microwave pulse jointly constitute the excitation scheme for the cascade of five photons. By scanning the detuning of the microwave, when the detuning of the microwave is exactly equal to the binding energy of the molecule, a Rydberg macrodimer can be formed. Figure~\ref{Fig14}(c) shows the calculated potential energy curves, and microwave photoassociation spectroscopy. Considering all molecular state of $M=0\sim6$ and symmetry of odd/even (-/+), Figure~\ref{Fig14}(c1) presents the adiabatic potential energy curve of the molecular state that can bind fine-structure-mixed macrodimers, which come from the strong dipolar flip ($D_{5/2}\leftrightarrow F_{5/2}$ and $D_{5/2}\leftrightarrow F_{7/2}$) and cross ($D_{5/2}F_{5/2}\leftrightarrow D_{5/2}F_{7/2}$) couplings. The zero-detuning in the figure represents the resonant transition frequency of $50D_{5/2}\to 48F_{5/2}$. Due to the inversion of the $nF_J$ Rydberg fine structure of cesium atoms, the transition energy of $50D_{5/2}\to 48F_{7/2}$ is lower. It can be clearly seen that near the nuclear distance of $R \approx 3~\mu$m, there exists a molecular binding potential well of the order of MHz. The measured high-resolution microwave spectroscopy is shown in Figure~\ref{Fig14}(c2). The narrow linewidth spectrum at the zero-detuning is the microwave spectrum of $50D_{5/2}\to 48F_{5/2}$ transition. The spectral line with a stronger signal and wider linewidth at approximately $2\pi\times$-20~MHz is the microwave resonance transition spectrum of the $50D_{5/2}\to 48F_{7/2}$ transition. Due to the dipole interaction of Rydberg atoms, the resonance spectral line of the $nF_{7/2}$ atom has a larger broadening, approximately 5 times that of the $nF_{5/2}$ atomic line. At the microwave detuning of $2\pi\times-10$~MHz, the microwave spectrum shows a relatively narrow molecular signal. The dashed line shows the comparison between theory and experiment. Figure~\ref{Fig14}(c3) shows the microwave optical association spectra simulated under two Wigner-Seitz radius conditions. The numerical simulation of the microwave photoassociation spectra accurately reproduces the Rydberg macrodimers characteristic spectral signals shown in Figure~\ref{Fig14}(c2).

\begin{figure}[htbp]
\begin{center}
\includegraphics[width=0.7\textwidth]{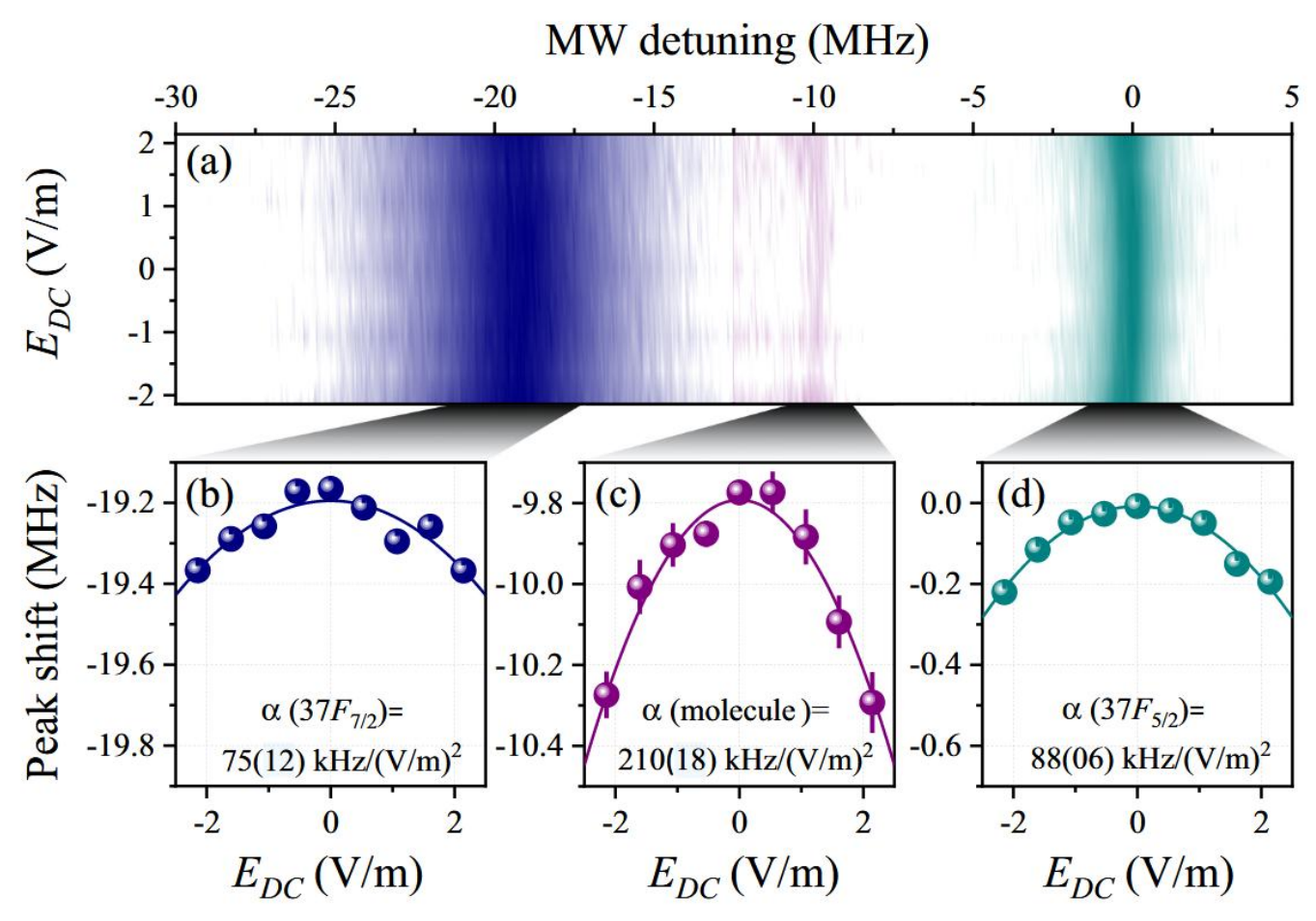}
\end{center}
\caption{ (a) Color-scale plot of microwave spectra near $37F_J$ vs DC electric field $E_{DC}$ and microwave detuning from the atomic $39D_{5/2} \to 37F_{5/2}$ transition. Three enlarged subplots show the Stark shifts of the (b) $37F_{7/2}$ atomic line (blue), (c) the $39D_{5/2}-37F_{J}$ Rydberg macrodimer (purple), and (d) the $37F_{5/2}$ atomic line (cyan). The solid lines show the best-fit functions of $W=-\frac{1}{2}\alpha E^2_{DC}+\beta$, with the fitted polarizabilities $\alpha$ shown in the panels~\cite{baiprr2024}. (Reproduced with permission and adapted from Ref.~\cite{baiprr2024}, licensed under a Creative Commons Attribution 4.0 International license.)
}\label{Fig15}
\end{figure}

The formation of this new type of Rydberg macrodimers is universal and can cover a wide range of principal quantum numbers. The extracted molecular binding energy for the principal quantum number range $n=39-48$ scalings as $n_{\text{eff}}^{-2.90(8)}$. Furthermore, they found that this kind of molecule is more sensitive to the external electric field. Figure~\ref{Fig15}(a) displays the microwave photoassociation spectroscopy under different DC  electric fields. By measuring the Stark frequency shifts of atomic and molecular signals, as shown in Fig.~\ref{Fig15}(b-d), it was found that the polarizability of the molecules is approximately 2.5 times that of the atoms, which indicates that molecules are more sensitive to external weak electric fields compared to atoms. This work not only reveals a new binding mechanism for the Rydberg macrodimers, extending their quantum states to high angular momentum $F$ states, but also provides new experimental evidence for the induced dipole interaction within the molecules.

In addition to alkali metal atoms, such as Rb and Cs, alkaline earth-like elements~\cite{Wojciechowska2025}, noble gases~\cite{Verdegay2025}, and strange molecular elements~\cite {Peper2021,Rayment2024} can also form exotic Rydberg macrodimers through bound potential wells near avoided crossings in asymptotic potential energy curves. In summary, Rydberg macrodimers, bound by multipole long-range interactions between two or more highly excited Rydberg atoms, have attracted considerable attention in recent years in both theoretical and experimental research. The investigation of Rydberg macrodimers has expanded our understanding of the coherent dynamics~\cite{Magoni2023} of  ultracold atomic systems, the measurement of correlations of quantum gases~\cite{stecker2017}, and aspects such as vacuum fluctuations~\cite{Ford2011Effects,Menezes2015Vacuum}.

\section{Ion-Rydberg molecules}\label{sec4}

An ion-Rydberg molecule composed of one ion and one Rydberg atom, is another kind of Rydberg molecule, bound with the electric single or multipolar interactions between Rydberg atoms and ions. Duspayev et al. first theoretically calculated the molecular potential energy curves and vibrational energy levels, and then proposed an experimental scheme for Cs and Rb atoms~\cite{Duspayev2021Long,Duspayev2022}. Meanwhile, Denschlag et al. also presented a theoretical model of ion-Rydberg molecules and calculated the potential energy curve of Rydberg molecules~\cite{Dei2021}.

\begin{figure}[htbp]
\begin{center}
\includegraphics[width=0.9 \textwidth]{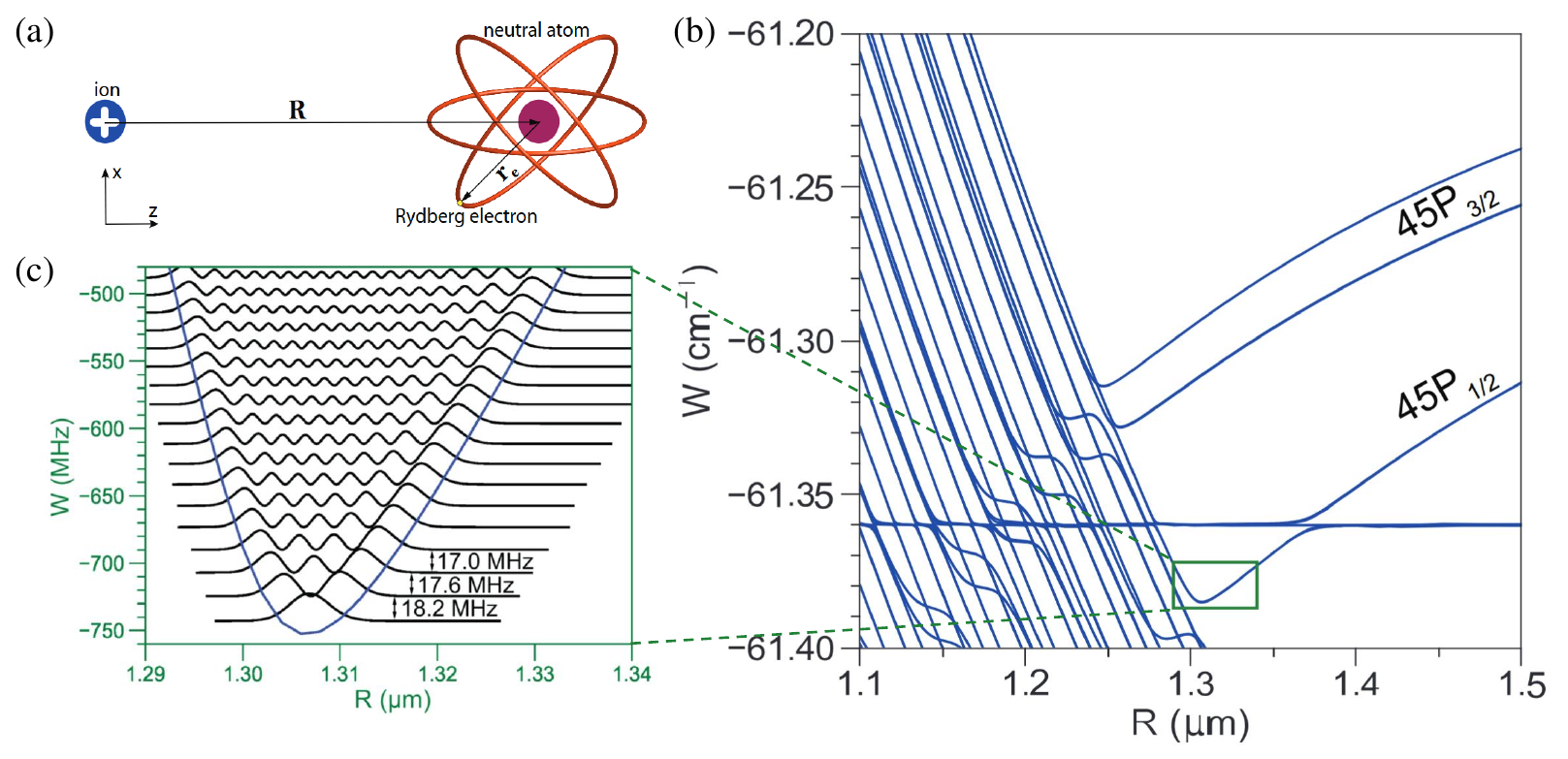}
\end{center}
\caption{ (a) Ion-Rydberg molecule schematic diagram. The molecular quantization axis and the internuclear distance $\textbf{R}$ between the ion and the neutral atom are both along the $z$-axis. The relative position of the Rydberg electron in the neutral atom is $r_e$. (b) Ion-Rydberg molecule potential energy curves as a function of internuclear distance $\textbf{R}$ relative to the atomic ionization potentials for rubidium $45P_J$ state. (c) A magnified view of the minimum of the lowest potential energy curve in the green square in (b). The lowest 18 vibrational states in this PEC well are shown in Ref.~\cite{Duspayev2021Long}. (Reproduced with permission and adapted from Ref.~\cite{Duspayev2021Long}, licensed under a Creative Commons Attribution 4.0 International license.)
}\label{Fig16}
\end{figure}

A schematic diagram of the ion-Rydberg molecule is shown in Figure~\ref{Fig16}(a), where  $\textbf{R}$ represents the nuclear distance between the free ion and the neutral Rydberg atom, and $r_e$ represents the distance between the Rydberg electron and the Rydberg atomic nucleus. The total Hamiltonian of the system is expressed as:
\begin{equation}
\hat{H} = \hat{H}_0 + \hat{V}_{\mathrm{I}},\label{eq9}
\end{equation}
where, $\hat{H}_0$ represents the Hamiltonian of the Rydberg atom, and the interaction $\hat{V}_{\mathrm{I}}$ between the ion and the Rydberg atom is:
\begin{equation}
\hat{V}_I
= - \frac{e^2}{4\pi\varepsilon_0}\sum_{K=1}^{\infty}\sqrt{\frac{4\pi}{2K+1}}\frac{r_e^{K}}{R^{K+1}}Y_{K,0}(\theta_e,\phi_e),\label{eq10}
\end{equation}
where $Y_{K,0}(\theta_e,\phi_e)$ is spherical harmonics function, depending on the angular position of the Rydberg electron, and $K$ is the order of the multipole interaction. $\varepsilon_0$ is the vacuum dielectric constant, and $e$ is the elementary charge. When $K=0$, the monopole interaction of the neutral Rydberg atom vanishes. Considering the Born-Oppenheimer approximation and ignoring higher-order multipoles, we diagonalize the Hamiltonian of the system to obtain the potential energy of rubidium (cesium) ion-Rydberg molecules by using the non-perturbative Rydberg state as the basis vector. 

The potential energy curves for rubidium $45P_J$ ion-Rydberg molecules are shown in Figure~\ref{Fig16}(b). The order of the electrostatic multipole interaction is $K_{max}=6$. The atoms are initially in the initial state $5D_{3/2}$, and the photoassociation laser is assumed to be $\sigma^-$ polarized with a field amplitude of $1 \times 10^6$~V/m. The angle $\theta$ between $E_L$ (homogeneous electric field) and the molecular axis is $45^{\circ}$. It can be seen that at a distance of $1.3~\mu$m, the fine structure energy levels of the $P$-state Rydberg atom generate repulsive interactions, forming a bound potential well with the high-lying multiplets, which can bind free ions and form an ion-Rydberg molecule. Further theoretical research indicates that the molecular bound states are expected to be observable near $nP_J$ Rydberg states in both Rb and Cs, over a wide range of $n$, that is the general shapes of the potential energy curves neither depend on $n$ nor on type of atom. 

The potential wells within the PECs are typically several gigahertz deep, as illustrated in Figure~\ref{Fig16}(b), and can support numerous vibrational states. This feature makes ion-Rydberg molecules particularly well-suited for laser spectroscopic investigations and time-dependent wave-packet dynamics studies. Figure~\ref{Fig16}(c) presents an enlargement of the lower PEC well associated with the $45P_J$ state (indicated by the green square in Figure~\ref{Fig16}(b)), along with the squared vibrational wavefunctions for the lowest 18 vibrational levels. The vibrational spectrum resembles that of a perturbed harmonic oscillator, where the level spacings follow $f_n=f_0+\alpha_1\nu+\alpha_2\nu^2+\cdot\cdot\cdot$ with $f_0$ representing the fundamental vibrational spacing, $\nu= 0,1,2,\cdot\cdot\cdot$ denoting the vibrational quantum number. A fit to the lowest 10 vibrational levels shown in (c) yields $f_0 = 18.2$~MHz, $\alpha_1 = −0.61$~MHz, and $\alpha_2 = 0.02$~MHz for the first- and second-order dispersion coefficients, respectively. 

\begin{figure}[htbp]
\begin{center}
\includegraphics[width=0.95\textwidth]{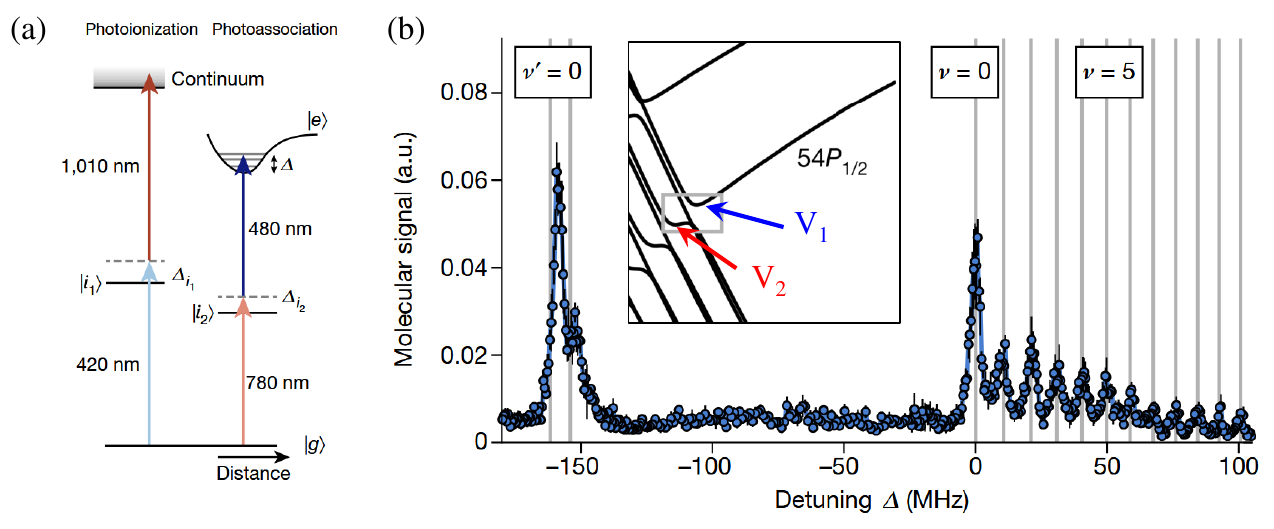}
\end{center}
\caption{ (a) Photoionization and photoassociation process used for molecule creation. 420~nm (with a detuning $\Delta i_1$) and 1010~nm lasers are used to produce ions. 780~nm (with a detuning $\Delta i_2$) and 480~nm lasers finish the preparation of Rydberg atoms by two-photo excitation. (b) Normalized molecular signal as a function of the Rydberg laser detuning $\Delta$, averaged over eight experimental cycles. In the detuning range from $−43$~MHz to $−122$~MHz, the step size increases from 0.5~MHz to 1~MHz. The gray lines mark the calculated vibrational states. Error bars represent the standard errors of the mean. a.u., arbitrary units. The illustration shows two potential wells $V_{1(2)}$ corresponding to the calculations~\cite{zuber2021}. (Reproduced with permission and adapted from Ref.~\cite{zuber2021}, licensed under a Creative Commons Attribution 4.0 International license.)}\label{Fig17}
\end{figure}

Later on, Zuber et al. reported the experimental observation of the ion-Rydberg molecular signal with a bond length of several micrometers~\cite{zuber2021}, where the vibrational spectrum and spatial resolution of the bond length and angular alignment of the molecule were reported using a high-resolution ion microscope. Figure~\ref{Fig17}(a) displays the experimental energy level scheme. Cold rubidium ions are generated through two-photon ionization of ground-state rubidium atoms. The  atoms in the ground state can be excited to the Rydberg state  via a two-photon process (780~nm + 480~nm) near the ion. Figure~\ref{Fig17}(b) shows the measured vibrational spectrum of the ion-Rydberg molecule corresponding to the $|54P, |m_j| = 1/2\rangle$ states at large distances. Well-separated peaks in the spectrum correspond to vibrational states, where the peak marked with $\nu=0$ is attributed to the lowest vibrational state of the outer molecular potential well $V_1$. The 11 smaller peaks on the blue-detuned side indicate higher vibrational levels of the molecule with a maximum vibrational spacing of approximately 11~MHz. The peaks marked with $\nu'=0$ on the red side are attributed to the lowest vibrational state of the potential well $V_2$. The energies of these vibrational levels are in good agreement with the theoretically predicted positions (gray lines). Besides, the measured lifetime of the ground vibrational state is $11.5 \pm 1.0 ~\mu\text{s}$, which is much shorter than the theoretically expected lifetime of the molecule.

The investigation of ion–Rydberg molecules is still at an early exploratory stage~\cite{Zou2023,Maran2025}. Further progress requires the refinement of the theoretical descriptions of the electrostatic multipole interactions between Rydberg atoms and ions, as well as the development of strategies to enhance the photo-association efficiency of ion–Rydberg molecular states. At the same time, a more systematic understanding of the underlying molecular structure and dynamics is essential to establish a solid foundation for ion–Rydberg molecular physics. On the experimental side, it is crucial to propose and implement feasible schemes for the controlled preparation of different classes of ion–Rydberg molecules, thereby enabling the realization of more stable and long-lived molecular states. Such advances would provide a versatile physical platform for applications in quantum information processing, ultraweak signal detection, and precision measurements. Moreover, ion–Rydberg molecules offer a promising testbed for investigating few-body and many-body quantum phenomena, owing to their long-range interactions and high tunability, opening new avenues for exploring complex quantum systems.

\section{Challenges and outlook}\label{sec5}

Rydberg molecules, often described as ``giant molecules" that bind highly excited Rydberg atoms with ground state atoms, Rydberg atoms or ions, have stood at the forefront of interdisciplinary research since their inception.  They are formed through unconventional bond mechanisms~\cite{Greene2000,Boisseau2002,Duspayev2021Long}  beyond the framework of traditional chemical bonds. Their significance extends far beyond the verification of a theoretical curiosity; rather, they fundamentally broaden our understanding of molecular formation and interactions, providing a ``super laboratory" for exploring long-range interactions, quantum correlations, and many-body physics.

From an applied perspective, Rydberg molecules have emerged as a shining ``quantum star": their intrinsically large electric dipole moments and strong interactions make them promising candidates for implementing quantum logic operations and constructing scalable quantum information processing architectures. At the same time, advances in Rydberg atoms have established them as powerful quantum simulators for complex condensed-matter models, such as spin ice~\cite{Samajdar2025} and topological order~\cite{Semeghini2021}, and as highly sensitive probes of microwave electric fields~\cite{jingAtomicSuperheterodyneReceiver2020}, opening a new era of quantum sensing. Importantly, recent experimental progress has demonstrated that Rydberg molecules can be prepared and controlled in programmable atomic arrays, opening the door to their use in quantum simulations in engineered lattice geometries~\cite{Guttridge2025}. Building on these developments, Rydberg molecules are emerging as a promising platform that extends these capabilities through their richer internal structures and interaction properties~\cite{overstreet2009,Hummel2021pra}.

As a bridge connecting the microscopic quantum world with macroscopic technological applications, Rydberg molecules continue to open new chapters in the search for novel matter and are poised to play an indispensable, engine-like role in the advancement of ultracold atom-based quantum technologies. Looking ahead, research in this field is evolving from foundational studies toward greater depth and complexity, and from isolated single quantum systems toward controllable ensembles. The community is entering a new stage focused on the synthesis and control of multi-atom and heteronuclear Rydberg molecular clusters~\cite{Zhang2022,Wang2022,Guttridge2023}, which is expected to unlock rich phases and novel quantum phenomena. Even more exciting is the prospect of integrating the exceptional properties of Rydberg molecules with emerging technologies such as topological photonics and integrated quantum photonic chips, which may allow topologically protected quantum information processing applications that could overcome bottlenecks in the realization of practical quantum technologies.
    
\addcontentsline{toc}{chapter}{Acknowledgment}

\section*{Acknowledgment}
J. Bai, Y. Jiao and J. Zhao thank the supported by the National Natural Science Foundation of China (No. 12241408, U2341211, 12120101004, and 12504304); Changjiang Scholars and Innovative Research Team in University of Ministry of Education of China (No. IRT$\_$17R70); and Shanxi Province's Basic Research Program (202503021212074). X. Q. Shao was supported by the National Natural Science Foundation (Grant.12174048). WL. acknowledges support from the EPSRC through Grant No. EP/W015641/1 and No. EP/W524402/1.

\renewcommand{\refname}{References}
%\bibliographystyle{unsrt}
%\bibliography{main}

\end{document}